\documentclass[12pt,osajnl2,preprint,showpacs,superscriptaddress]{revtex4}

\usepackage{graphicx}
\usepackage{amsmath,amssymb}

\begin{document}
\title{Nonlinear effects in random lasers}
\author{Jonathan Andreasen}
\affiliation{
Laboratoire de Physique de la Mati\`ere Condens\'ee, CNRS UMR 6622,
Universit\'e de Nice-Sophia Antipolis,
Parc Valrose, 06108, Nice Cedex 02, France}
\author{Patrick Sebbah}
\affiliation{
Laboratoire de Physique de la Mati\`ere Condens\'ee, CNRS UMR 6622,
Universit\'e de Nice-Sophia Antipolis,
Parc Valrose, 06108, Nice Cedex 02, France}
\affiliation{
Institut Langevin, ESPCI ParisTech, CNRS UMR 7587,
10 rue Vauquelin, 75231 Paris Cedex 05, France}
\author{Christian Vanneste}
\affiliation{
Laboratoire de Physique de la Mati\`ere Condens\'ee, CNRS UMR 6622,
Universit\'e de Nice-Sophia Antipolis,
Parc Valrose, 06108, Nice Cedex 02, France}

\date{\today}

\begin{abstract}
  Recent numerical and theoretical studies have demonstrated that the modes at threshold of a random laser are in direct correspondence with the resonances of the same system without gain, 
  a feature which is well known in a conventional laser but which was not known until recently for a random laser. 
  This paper presents numerical results, which extend such studies to the multimode regime that takes place when the pumping rate is progressively increased above threshold. 
  Behavior that is already known in standard lasers, such as mode competition and nonlinear wave-mixing, are shown to also take place in random lasers thus reinforcing their recent modal description. 
  However, due to the complexity of the laser modes and to the openness of such lasers, which requires large external pumping to compensate for strong loss, 
  one observes that these effects are more pronounced than in a conventional laser.
\end{abstract}


\maketitle

\section{Introduction}

Since their prediction by Lethokov \cite{Letokhov68a}, random lasers have been the subject of numerous studies.
Contrary to conventional lasers, they have no cavity like a Fabry-P\'erot resonator or a ring cavity.
Instead, they are made of a scattering random medium like a semiconductor powder or a suspension of scattering particles in a laser dye, and are excited by an external pump which introduces gain. 
Multiple scattering of light in the random medium provides optical feedback. 
An important feature of random lasers is that they are open systems, typically with strong leakage. 
Because of these unusual features, there was a several years' debate about the nature of random lasing. 
Random lasing has been described as light diffusion with gain \cite{John96,Wiersma97}, 
or in terms of well-localized modes inside the scattering medium \cite{Cao2000,Jiang2000,Vanneste2001,Polson2002,Alpakov2002}, 
or by random walks of photons along exceptionally long paths \cite{Mujumdar2004}. Such approaches suffered from different drawbacks. 
On the one hand, the diffusion equation approach and the random walk of photons do not take into account the wave aspect of light propagation, 
thus neglecting the part played by interference. 
On the other hand, experimental characterization of the scattering media which exhibited random lasing in the presence of gain showed that they were far from the localization regime, 
thus making the occurrence of well-localized modes unlikely.

Recently, important progress has been made in theoretical and numerical studies by recognizing that even the bad resonances of leaky systems play a role in random lasing, 
similar to the part played by cavity modes of a conventional laser \cite{Vanneste2007,Tureci2008,Tureci2009}. 
In particular, it was shown \cite{Andreasen2011} that the first lasing mode at threshold is not strictly identical but very close to a resonance, or quasimode, of the passive system. 
Despite the complex nature of these quasimodes, the understanding of random lasing is therefore greatly simplified in the single mode regime, just above threshold. 
As the pumping rate is increased, however, nonlinear effects come into play.
These effects might be even more pronounced in weakly scattering active media where high external pumping is required to compensate for strong loss. 
This is particularly true in the multimode regime above the lasing threshold, with the onset of competition between different lasing modes. 
Detailed investigations of the multimode regime in random lasers are scarce \cite{caoPRB03,jiangPRB04,conti08}. 
The background of standard laser physics is not directly transposable to random lasers, 
where the complexity of the spatial and spectral properties of random lasers need to be taken into account. 
The recent steady-state \textit{ab initio} laser theory (SALT) \cite{Tureci2006,geOE08}, 
which considers the openness of novel laser systems, such as random lasers, 
is the first theory to give predictions about nonlinear phenomena and mode interaction \cite{Tureci2008,Tureci2009}. 
This theory relies, however, on several assumptions. 
In particular, it assumes the existence of a steady-state multiperiodic solution of the laser field, population inversion and polarization of the atomic medium.
However, the full nonlinear dynamics certainly play a role in determining behavior in such complex and highly nonlinear media. 
Including them would not only bring a more complete theoretical approach, but also bring theory closer to experiment.

In this paper, we numerically investigate the full dynamics of one-dimensional (1D) and two-dimensional (2D) random lasers using steady external pumping by progressively increasing the external pump intensity. 
We report several manifestations of optical nonlinearities, both in the single mode and multimode regime, in the steady state as well as in the transient regime. 
Strong relaxation oscillations, mode competition and mode suppression are reported in the transient regime, which reveal the complexity of the laser dynamics in random lasers. 
Third-harmonic generation, four-wave mixing and sum-frequency generation are observed, which have never been reported before.
Above the so-called second threshold \cite{Roldan2005}, the steady state becomes unstable. 
A coherent instability manifests itself as temporal oscillations of the field intensity, atomic population inversion and medium polarization. 
This issue has been addressed in a different paper \cite{andreasenLH} and will not be discussed here, although it is another manifestation of the nonlinear dynamics expected in random lasers.
If these effects are not new, their observation in random lasers is interesting since they not only challenge the theoretical understanding of these systems but they are more pronounced than in conventional lasers. 
Our effort here is to stress how these observations are related to the particular nature of the modes of random lasers.

The paper is organized as follows.
Section \ref{sec:model} describes the 1D and 2D random structures that have been studied and the numerical methods that have been used.
Section \ref{sec:competition} shows nonlinear effects like the build-up of laser oscillation and interactions between lasing modes, 
which include mode competition and nonlinear wave-mixing.
Conclusions are given in Section \ref{sec:conclusion}.

\section{Numerical approaches\label{sec:model}}

\subsection{Random structures}

\begin{figure}
  \includegraphics[width=6.8cm]{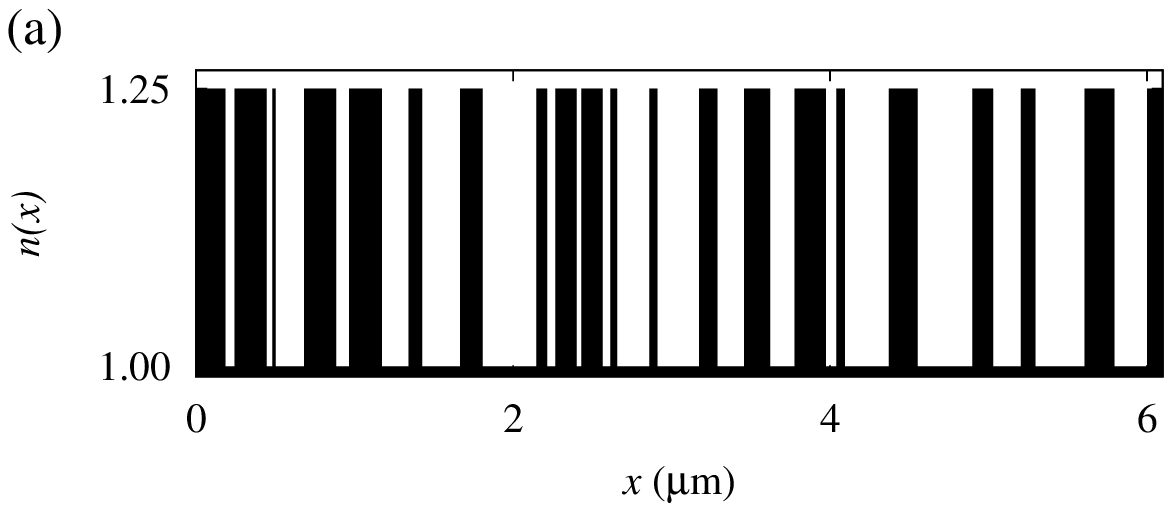}\\
  \includegraphics[width=6.0cm]{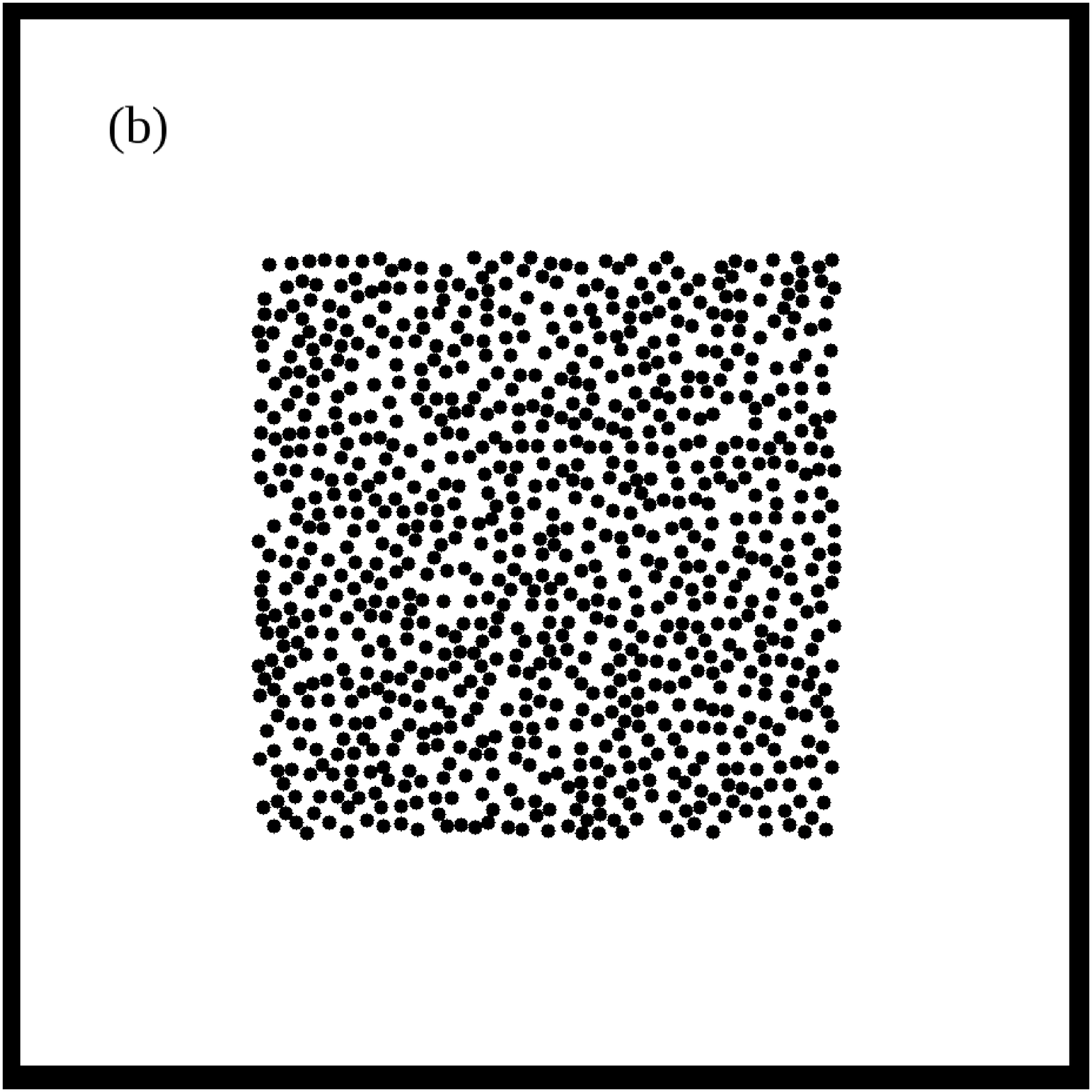}
  \caption{\label{fig:fig1}
    Spatially dependent index of refraction. 
    (a) $n(x)$ of a 1D random structure.
    (b) $n(x,y)$ of a 2D random structure, $5\times 5$ $\mu m^2$.
    The outer black border represents the absorbing boundary.
  }
\end{figure}

The 1D random structures we consider are similar to those studied in \cite{Andreasen2009,Andreasen2010}.
They are composed of $41$ layers and shown in Fig. \ref{fig:fig1}(a).
Dielectric material with optical index $n_1=1.25$ separated by air gaps
($n_2=1$) results in a spatially modulated index $n(x)$.
Outside the random medium, the index is 1.
The system is randomized by specifying thicknesses for each
layer as $d_{1,2} = \left<d_{1,2}\right>(1+\eta\zeta)$, where
$\left<d_1\right>=100$ nm and $\left<d_2\right>=200$ nm are the average
thicknesses of the layers, $\eta = 0.9$ represents the degree of randomness,
and $\zeta$ is a random number in (-1,1).
The length of the random structure $L$ is normalized to $\left<L\right>=6.1$ $\mu$m.
These parameters give a localization length $\xi \approx 11$ $\mu$m within the wavelength range considered here.

The 2D random structures considered [Fig. \ref{fig:fig1}(b)] are made of circular dielectric particles with radius $r=60$ nm,
optical index $n_1=1.25$, and surface filling fraction $\Phi = 40$\%, which are randomly distributed in a background medium of size $L^2=5\times 5$ $\mu$m$^2$ and index $n_2=1$. 
The optical index of the domain that surrounds the random medium is $n=1$.
The scattering mean free path $\ell_s\approx 2$ $\mu$m and the localization length $\xi \approx 12$ $\mu$m.
Such systems, which are similar to those considered in \cite{Sebbah2002,Vanneste2007}, are in a weakly scattering regime close to the diffusive regime.
We also consider a second set of smaller 2D systems with a size $L^2=1\times 1$ $\mu$m$^2$
having a density of states comparable to the 1D systems.
The optical index of the particles is chosen as $n_1=1.5$.
In this case, the scattering mean free path $\ell_s\approx 0.3$ $\mu$m and the localization length $\xi \approx 5$ $\mu$m.
Such small 2D systems are in a quasi-ballistic regime.

\subsection{Methods to study lasing\label{ssc:lmethods}}

In 1D and 2D systems the background medium of index 1 has been chosen as the active part of the system.
Two different methods have been used to study lasing in 1D systems. 
First, it is possible to employ the transfer matrix (TM) method similar to that used in \cite{Andreasen2011b}. 
However, this is a frequency-domain method, which cannot describe transient behavior in the time domain and mode competition in the multimode regime. 
For this reason, we also solve Maxwell's equations using the finite-difference time-domain (FDTD) method \cite{Taflove2005}. 
The two methods can be compared easily \cite{Andreasen2009}, yet bring different advantages.
In this case, the gain medium in FDTD is modeled by a four-level atomic system \cite{nagra98,Jiang2000,Sebbah2002}.

In 2D systems, we only use the FDTD method to solve Maxwell's equations and consider the case of transverse magnetic polarization. 
We use perfectly matched layer (CPML) absorbing boundaries \cite{Taflove2005} to approximate open boundary conditions. 
Gain in the system is again introduced by coupling Maxwell's equations with the population equations of a four-level atomic system. 
The corresponding equations are given in Appendix \ref{app:mectpe}. 
Population inversion between the levels corresponding to the laser transition is created by an external pump which transfers atoms from the 
ground state (level 1) to the upper level (level 4) of the four-level atomic system.
In the calculations presented in the paper, the control parameter is the pumping rate $P_r$ of atoms from the ground state to the upper level.

\section{Results\label{sec:competition}}

Our results here are devoted to random laser modes above threshold and to their interactions in the multimode regime.
We not only discuss the final stationary state but also the transient build-up of the laser field. 
In both situations, we observe noticeable mode competition. 
The large gain which must be introduced in order to counterbalance the large loss of the system leads to a significant evolution of the lasing modes as a function of the pumping rate. 
This effect is much larger than its counterpart in conventional lasers.
Finally, mode interaction reveals itself by the systematic observation of nonlinear wave-mixing,
which appears as soon as multiple lasing modes coexist.

\subsection{Laser field build-up}

Each FDTD calculation of a given random system with a fixed value of the pumping rate $P_r$ starts in the same way.
First, the initial atomic populations are set at the stationary values
in absence of stimulated emission, i.e., for $({E_z}/{\hbar\omega_a}){dP}/{dt}=0$ (Appendix \ref{app:mectpe}).
At the initial time $t=0$, a small-intensity broadband light pulse is launched in the system.
Since there is no spontaneous emission in our model, this short pulse, which propagates and is scattered by the particles of the system,
provides the non-zero initial field necessary for lasing action to begin.
If $P_r$ is larger than the threshold value,
the laser field builds up in the system until it reaches a value where the population inversion starts to be depleted by increasing stimulated emission.
The steady-state regime is reached when the value of the population inversion density corresponds to the gain,
which exactly compensates the loss through the open boundaries of the system.
This is the standard behavior of any laser oscillator.

\begin{figure}
\begin{center}
  \includegraphics[height=4.2cm,angle=270]{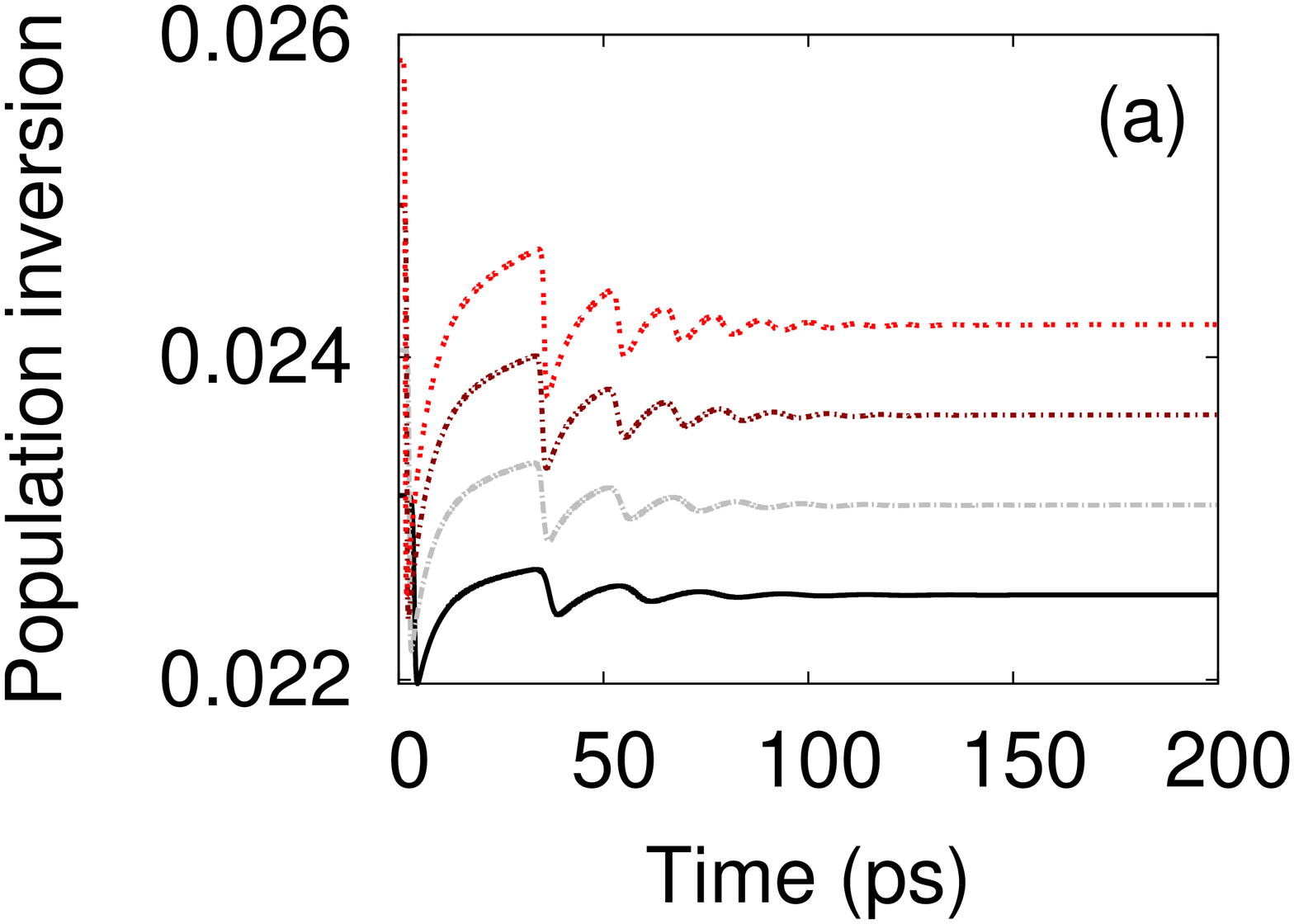}
  \includegraphics[height=4.2cm,angle=270]{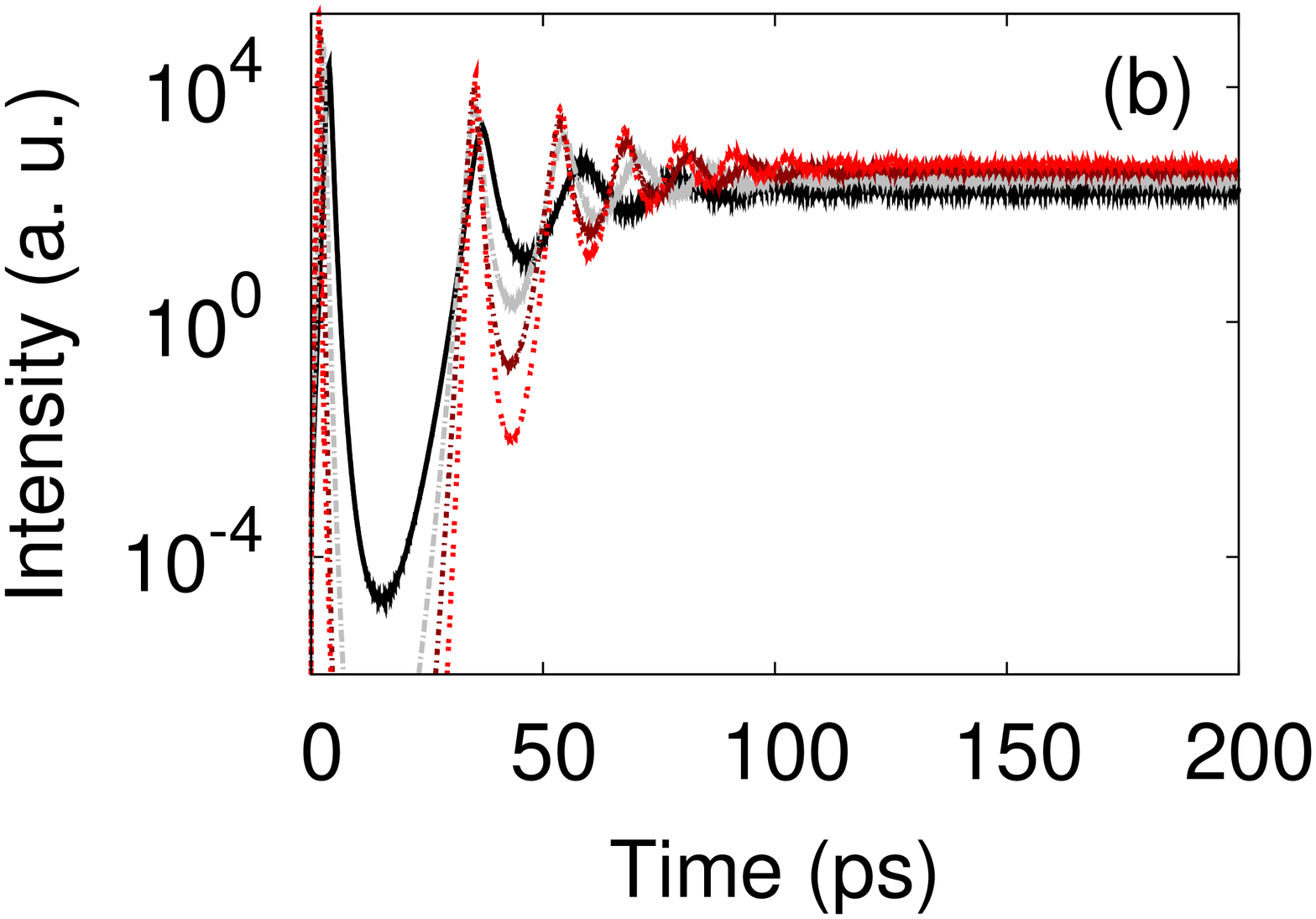}
  \caption{\label{fig:fig2} (Color online)
    Relaxation oscillations in (a) population inversion and (b) intensity,
    for $P_r=0.25$ (solid black line) to $P_r=0.28$ (dotted red line) where single-mode lasing occurs.
    The intensity and oscillation frequency increases with $P_r$.
  }
\end{center}
\end{figure}

However, random lasers are open systems usually with short ``cavity'' lifetimes, i.e., shorter than the atomic lifetime  $T_1$. 
This is the well-known condition \cite{Siegman1986} for the occurrence of relaxation oscillations or spiking, i.e., 
temporal oscillations of the field and of the population inversion before the laser oscillation converges to a steady state.
An example is given in Figure \ref{fig:fig2} where relaxation oscillations are seen in the population inversion and laser intensity for a 1D random laser. 
As the pumping rate increases, the intensity and the frequency of the relaxation oscillations increase. 
Indeed, relaxation oscillations in random lasers have been reported by several groups \cite{soukoulisPRB02,Noginov2004,Molen2010} 
and we have systematically observed them in our simulations thus stressing the importance of the openness of such systems.

\subsection{Mode alterations near threshold}

As mentioned earlier, even the bad resonances of open random systems play a role in lasing similar to the part played by the cavity 
modes of a conventional laser. 
In particular, the first lasing mode at threshold is not strictly identical but very close to a resonance or a quasimode of the 
passive system, a result which was established only recently for random lasers \cite{Andreasen2011}. 
However, this conclusion relies on several conditions that are not always fulfilled in real experiments,
for instance, uniform pumping of the total system.
In real random lasers, pumping takes place in the active elements, which usually do not fill up the total volume, 
such as particles suspended in laser dye where there is no gain in the particles.
This situation contrasts strongly with that of conventional lasers,
where pumping is usually distributed uniformly in the gain domain.

\begin{figure}
  \includegraphics[width=4cm,angle=270]{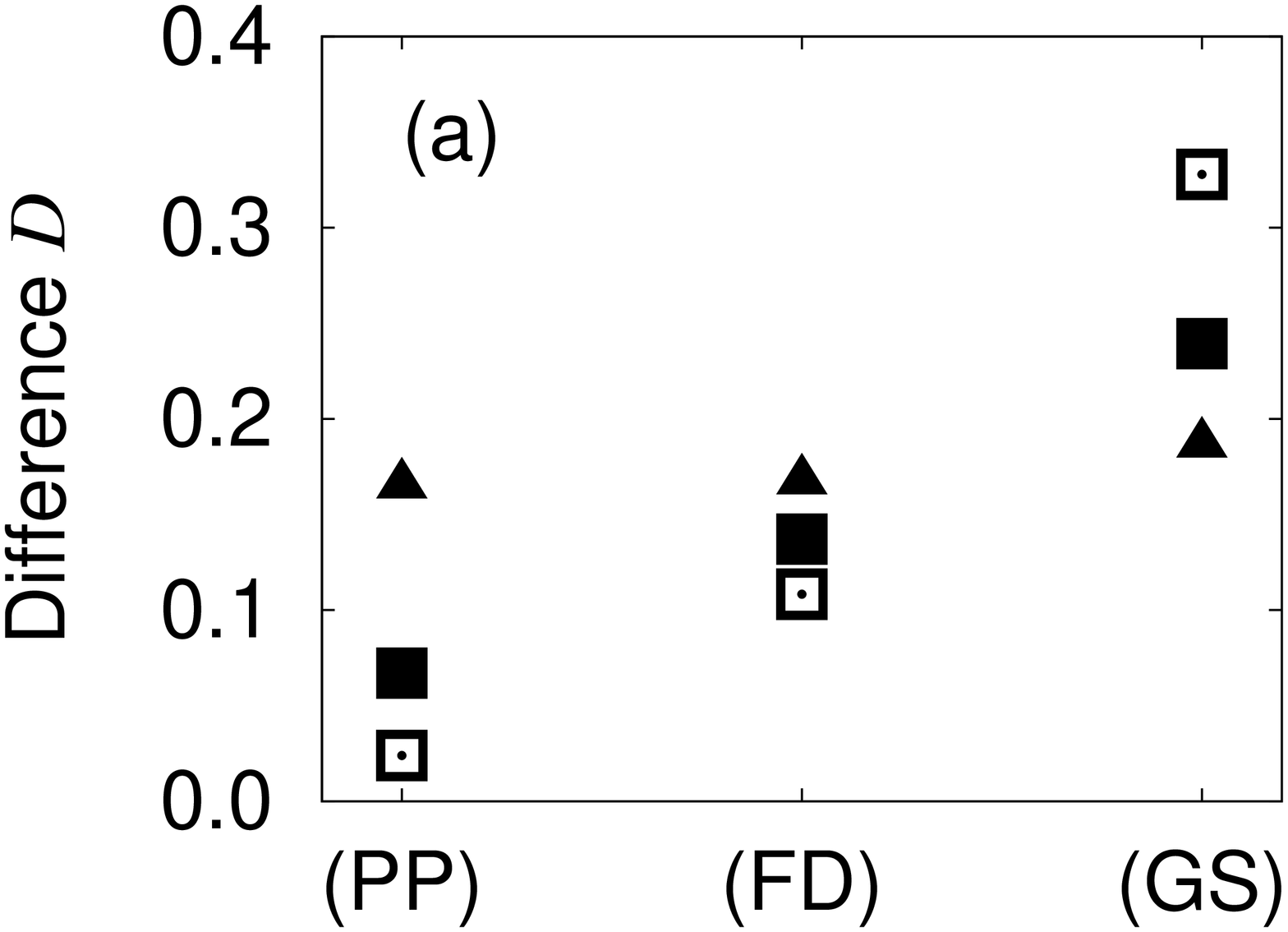}\\
  \includegraphics[width=4cm,angle=270]{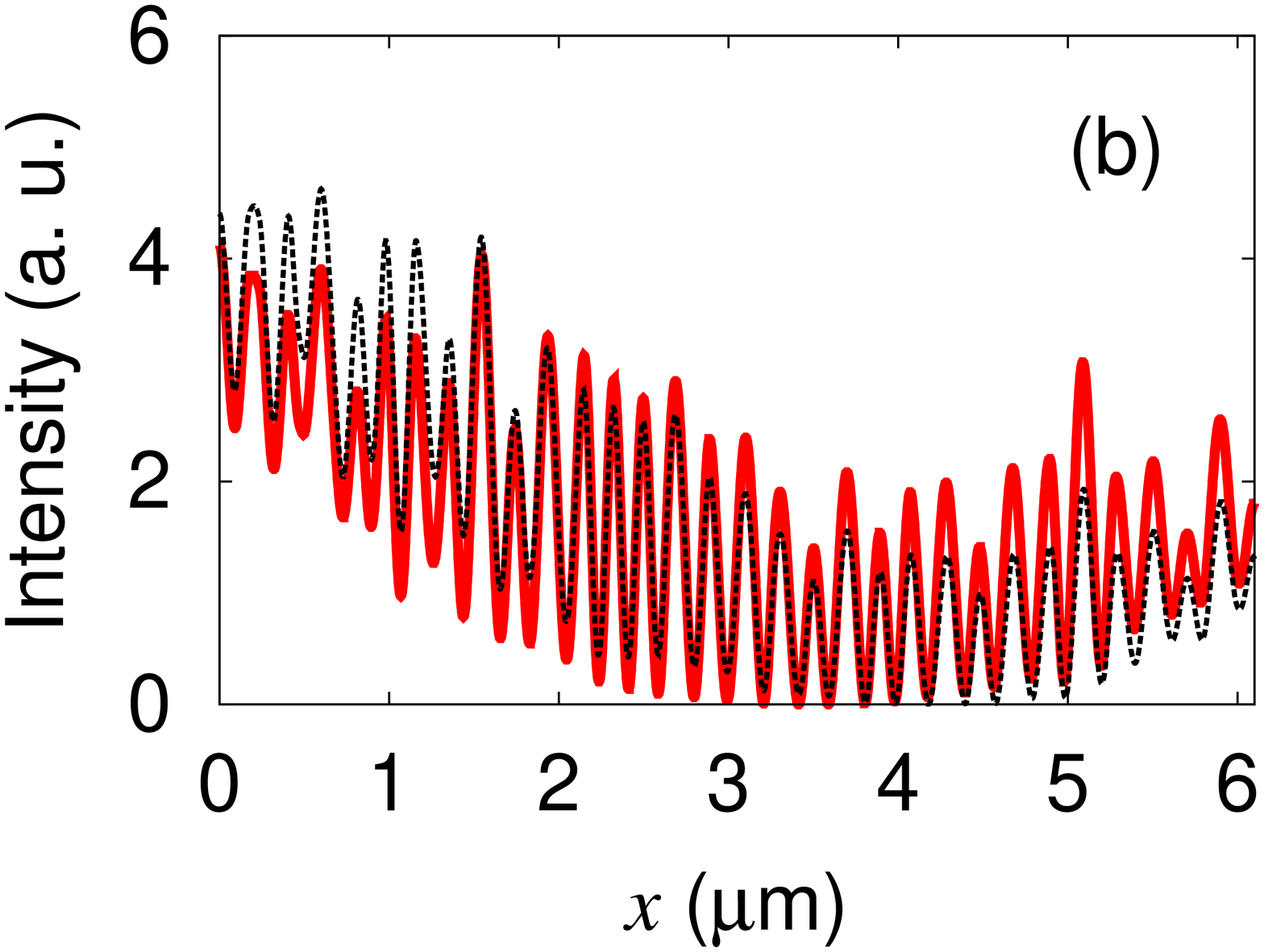}
  \caption{\label{fig:fig3} (Color online)
    (a) Differences between the intensity distributions of the threshold lasing mode with uniform and flat gain and
    that with successively including (PP) partial gain, (FD) frequency-dependent gain, and (GS) gain saturation.
    (triangles) first lasing mode, (squares) second lasing mode. 
    (open and closed squares) Two different realizations for second lasing modes to verify the effect of gain saturation.
    (b) Intensity distributions of second threshold lasing mode with (solid red line) uniform gain and (dashed black line) 
    gain saturation with gain only in the air gaps.
  }
\end{figure}

Three modes of a random laser near threshold are considered in Fig. \ref{fig:fig3}. 
One mode is the first lasing mode at threshold and the two other modes are second lasing modes above threshold for two different realizations of the disorder.
In order to obtain a quantitative measure of the similarity of two intensity distributions $E_i^2(x)$ and $E_j^2(x)$,
we introduce the difference
\begin{equation}
  \label{Difference}
  D_{ij}=\int_0^L\mid E_i^2(x)-E_j^2(x)\mid dx,
\end{equation}
where the distributions $E_{i,j}^2(x)$  are normalized to one.
The first column in Fig. \ref{fig:fig3}(a) compares the intensity distributions of the lasing modes 
with uniform gain and those with partial gain, i.e., gain located in the gaps between the high index particles [see Fig. \ref{fig:fig1}(a)].
Noticeable differences are clearly visible, up to the value $D=0.16$ for the first lasing mode.
Previous investigation of partial pumping in random lasers \cite{Andreasen2010b}
has already demonstrated that lasing modes can become very different from quasimodes of the passive system.
Here, we are in an intermediate situation where gain is not placed everywhere but homogeneously distributed across the system.
The impact of partial gain varies from mode to mode, as shown by the reduced $D$ values for the second lasing modes.
This illustrates the role of randomness in these lasers and the effects of partial pumping that arises in such multiple scattering systems.

Another approximation used in several theoretical and numerical studies is a flat gain curve, i.e., gain that does not depend on frequency. 
The second column in Fig. \ref{fig:fig3}(a) compares the intensity distributions of the first lasing mode with uniform and flat gain
and that with partial and frequency-dependent gain.
The frequency dependence of the gain does not contribute significantly to the modification of the spatial distribution of the first lasing mode,
which is close to the atomic transition wavelength, but does impact the second lasing modes.
The threshold for the second lasing mode is larger,
which results in a larger modification of the medium susceptibility due to the gain and further modifies the lasing mode.

The large alteration of the medium susceptibility due to the intense pumping that is required to achieve lasing is again a particular feature of random lasers due to their openness and associated strong leakage.
For comparison, we consider a Fabry-P{\'e}rot laser with mirrors of reflectivity 80\%.
The length between the mirrors is $L=6.1$ $\mu$m, the same as the random lasers. 
The effective index of the random lasers, $n_{eff}=1.1$, is used between the mirrors to obtain similar mode spacing.
The corresponding differences for this Fabry-P{\'e}rot laser are $D=10^{-4}$ and $10^{-3}$ for the first and second
lasing modes, respectively, compared to the smallest value $D=10^{-1}$ observed in the random laser.
This illustrates that although similar effects may be observed in random and conventional lasers, 
the inhomogeneity and the openness of random lasers make these effects more pronounced, causing more dramatic changes to steady-state laser behavior.

Finally, different from the first lasing mode at threshold, lasing modes in the multimode regime compete for gain with the occurrence of saturation effects.
Here, we take advantage of the TM and FDTD methods described in Sec. \ref{ssc:lmethods}. 
The third column in Fig. \ref{fig:fig3}(a) compares the intensity distributions of the first lasing mode with uniform and flat gain
and that with gain saturation computed via the FDTD method.
As expected, saturation effects at the threshold of the first lasing mode are negligible, but strongly influence the second lasing mode.
Figure \ref{fig:fig3}(b) compares the spatial profiles of the second lasing mode at threshold for these two situations.
The openness of the structure is demonstrated by the fact that the mode energy is peaked at the boundary of the system 
\cite{Wu2006,Vanneste2009,Andreasen2010b}.
For comparison, the corresponding differences for the Fabry-P{\'e}rot laser are $D=10^{-2}$ and $3\cdot 10^{-2}$ for the first and second lasing modes, respectively. 
Once again such effects in random lasers are more pronounced compared to their conventional laser counterpart.

\begin{figure}
\begin{center}
  \includegraphics[height=4.2cm,angle=270]{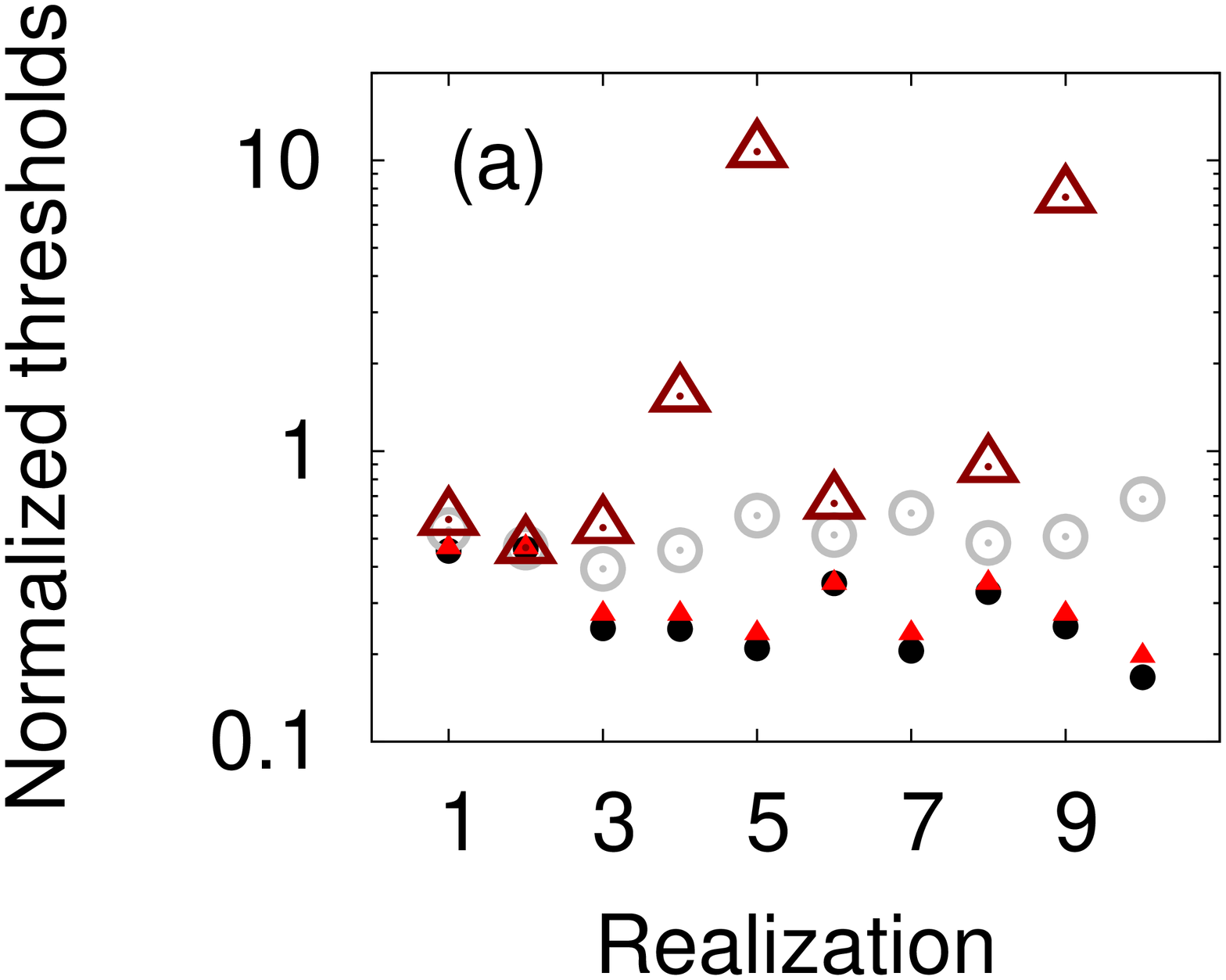}
  \includegraphics[height=4.2cm,angle=270]{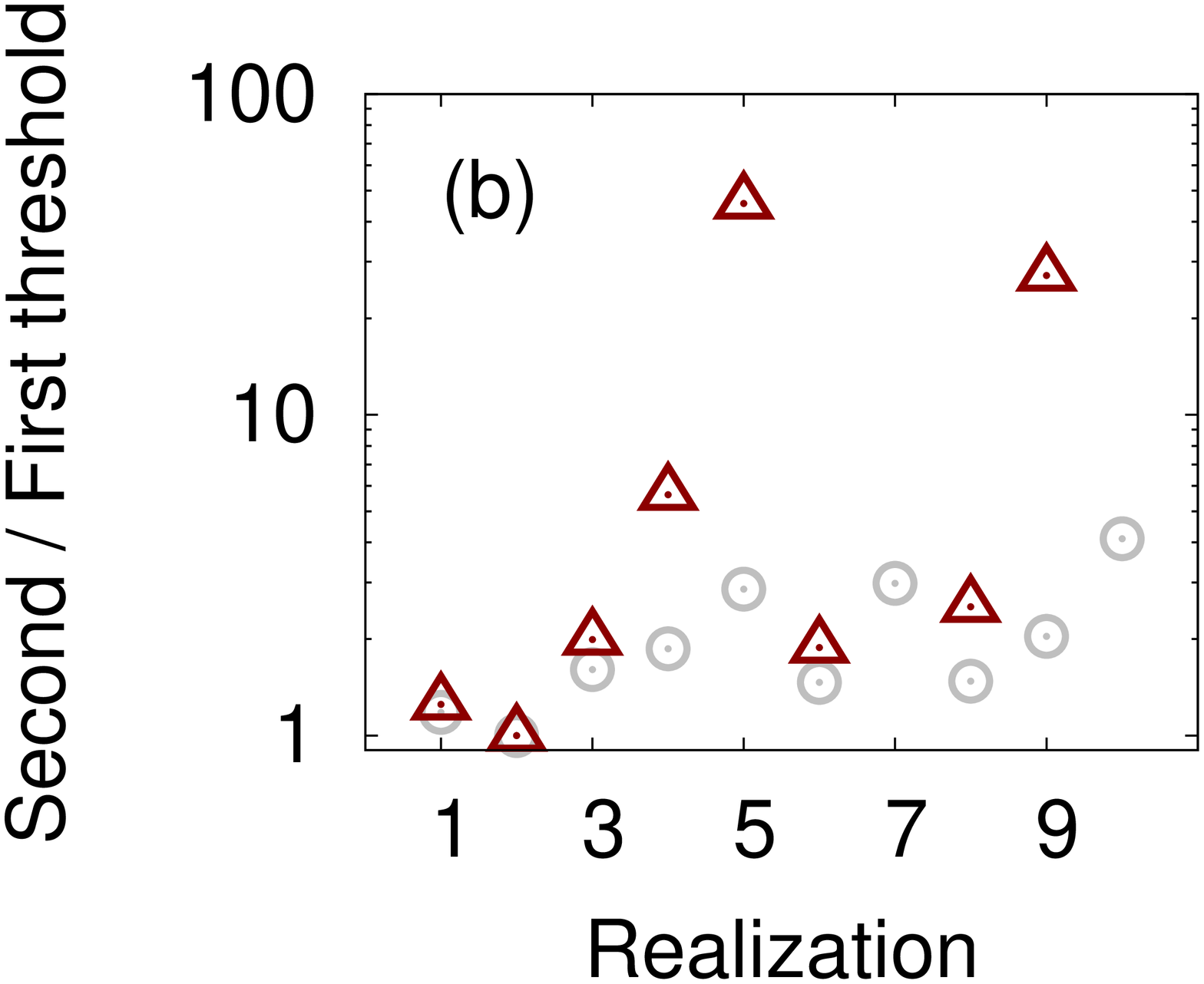}
  \caption{\label{fig:fig4} (Color online)
    Lasing thresholds for 10 realizations of 1D random lasers.
    (a) First (filled symbols) and second (open symbols) lasing thresholds without mode interaction (TM--circles) 
    and with mode interaction (FDTD--triangles) included.
    (b) Ratio of the second lasing threshold over the first lasing threshold without and with mode interaction included. 
    The single mode regime persisted for realizations 7 and 10, even for the largest ratio checked (9000).
  }
\end{center}
\end{figure}

In addition to the example case above,
we have performed a systematic investigation of lasing thresholds over ten different realizations of disorder in 1D random lasers. 
We compare results from the TM and FDTD methods, where the essential difference is that mode competition effects are included in FDTD.
Figure \ref{fig:fig4}(a) displays the lasing thresholds of the first and second lasing modes of the 10 random lasers. 
The lasing thresholds for the first mode are identical, whatever the method. 
This result is expected since, at the threshold of the first lasing mode, there is neither competition with other modes nor saturation effects. 
To the contrary, there is a noticeable difference between the two methods for the thresholds of the second lasing modes. 
This indicates that competition effects are important. 
The threshold value calculated via the FDTD method is always larger than the value calculated via the TM method. 
This demonstrates that the first lasing mode delays the onset of the second lasing mode, an effect that the TM method does not incorporate. 
One consistently notices that the larger the threshold difference calculated via the TM method between mode 1 and 2 
[filled and open circles in Fig. \ref{fig:fig4}(a)], 
the larger the difference for mode 2 between the threshold calculated via the FDTD method and that calculated via the TM method 
[open triangles and circles in Fig. \ref{fig:fig4}(a)]. 
This effect is clearly illustrated in Fig. \ref{fig:fig4}(b), which displays the ratio of the second lasing threshold over the first lasing 
threshold without (TM method) and with (FDTD method) mode interaction included. 
The larger the first ratio, even larger is the second ratio. 
Realizations 7 and 10 exhibit the largest threshold ratios via the TM method.
The second lasing thresholds were never reached with the FDTD method, even when the pumping rate was extended almost two orders of magnitude 
past the second lasing thresholds of other realizations.
In these systems, the single mode regime is very robust.
The quasimodes underlying the lasing mode in realizations 7 and 10 are examined in Appendix \ref{app:qmtlm}.

\subsection{Mode evolution above threshold}

Further above the lasing threshold, mode competition results in an irregular increase of the intensities of the lasing modes 
as function of the pumping rate and, possibly, in mode suppression. 
We confirm these predictions here. 
It is important to point out that in all cases, increasing the pumping rate eventually leads all the random lasers we have studied to reach the threshold of a 
coherent instability (called the second threshold) thus limiting the range of pumping rates for observing the stationary states we describe below. 
This is not surprising since such instabilities are known to occur in very leaky lasers \cite{Roldan2005}.
Details concerning such instabilities in random lasers have been discussed \cite{andreasenLH}, but their effect on mode interactions are beyond the scope of this paper. 
Thus, we focus now on laser behavior between the first threshold and the so-called second threshold.

\begin{figure}
\begin{center}
  \includegraphics[height=4.2cm,angle=270]{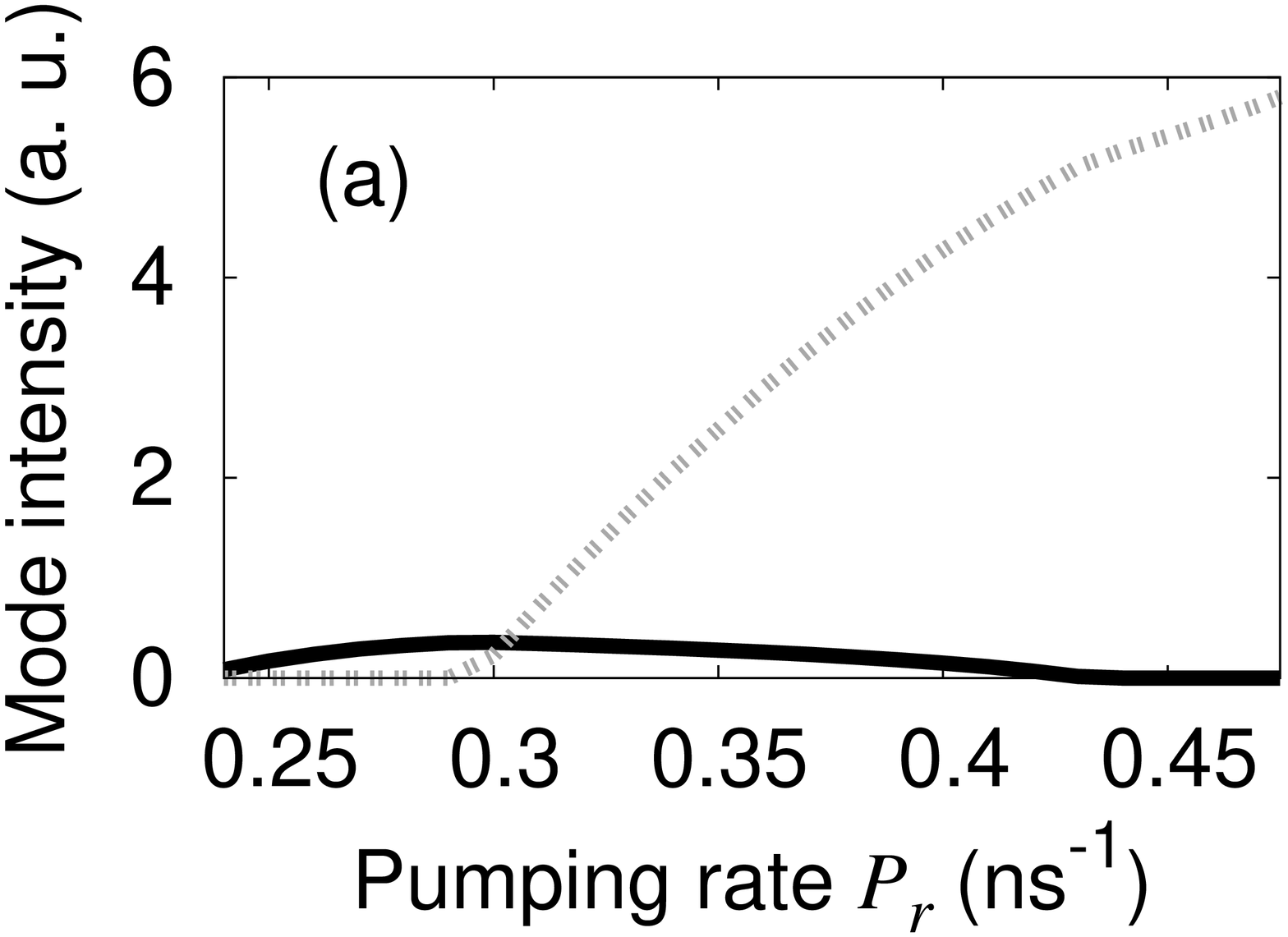}
  \includegraphics[height=4.2cm,angle=270]{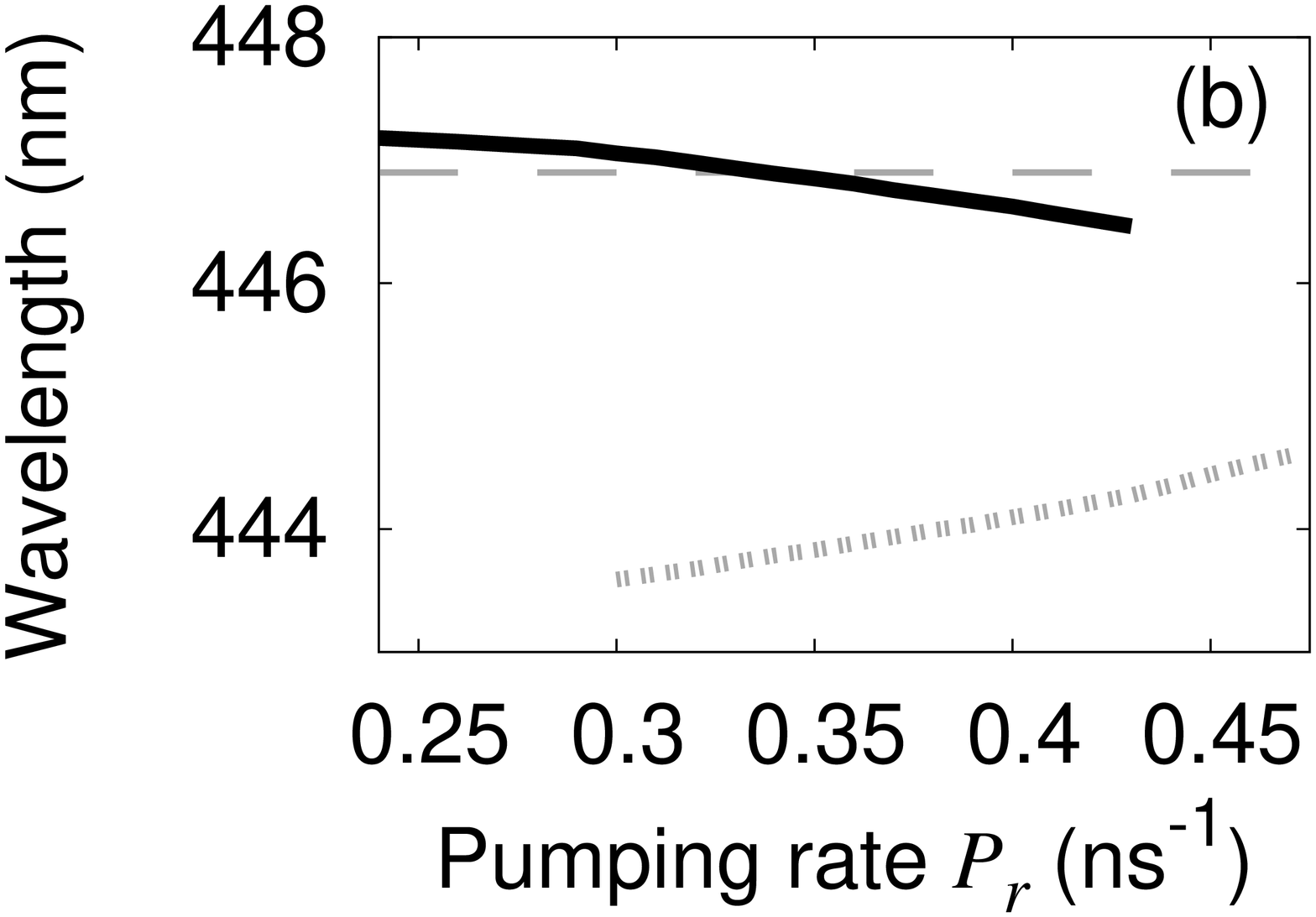}
  \caption{\label{fig:fig5}
    (a) Intensity and (b) wavelength vs. pumping rate $P_r$ of the (solid black line) first and (dotted gray line) second lasing mode. 
    Their respective thresholds are at $P_r=0.24$ and $P_r=0.30$. 
    Mode 1 is suppressed for $P_r \ge 0.44$. 
}
\end{center}
\end{figure}

Figure \ref{fig:fig5} shows an interesting case of a bimodal 1D random laser emission spectrum evolution as the pump rate is progressively increased. 
As the second lasing mode turns on at $P_r=0.30$, 
the progression of the first mode is stopped and its intensity progressively reduces until it turns off at $P_r=0.44$ [Fig.~\ref{fig:fig5}(a)]. 
Figure \ref{fig:fig5}(b) displays the wavelengths of the two lasing modes as a function of the pumping rate. 
Frequency drift of mode 2 toward the central frequency of the gain curve is a manifestation of frequency pulling
and allows the mode to experience more amplification, which certainly helps mode 2 to overcome mode 1 \cite{Note1}.
However, the main reason for the extinction of the first lasing mode is self-saturation and cross-saturation effects between the two modes in agreement with the recent predictions of SALT \cite{Tureci2008}.

\begin{figure}
\begin{center}
  \includegraphics[height=1.9cm]{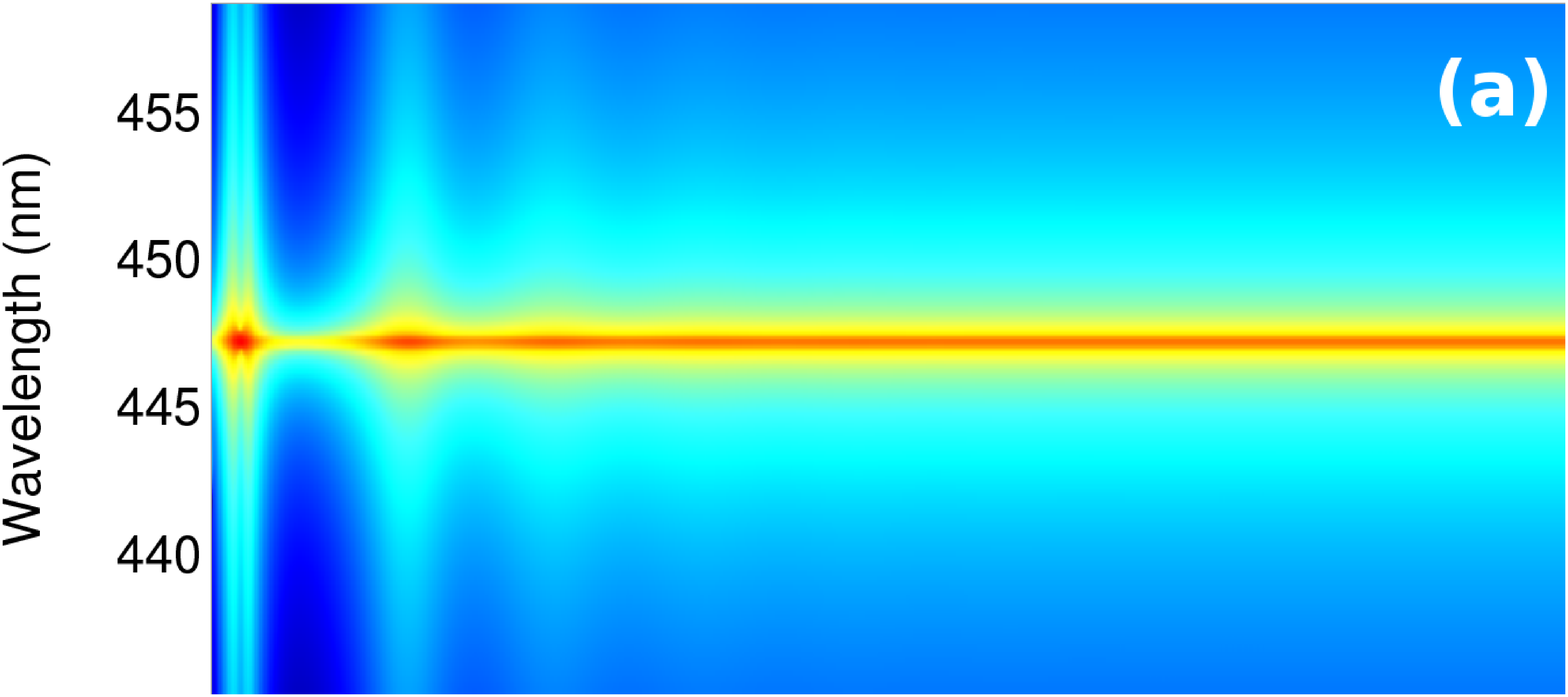} 
  \includegraphics[height=1.9cm]{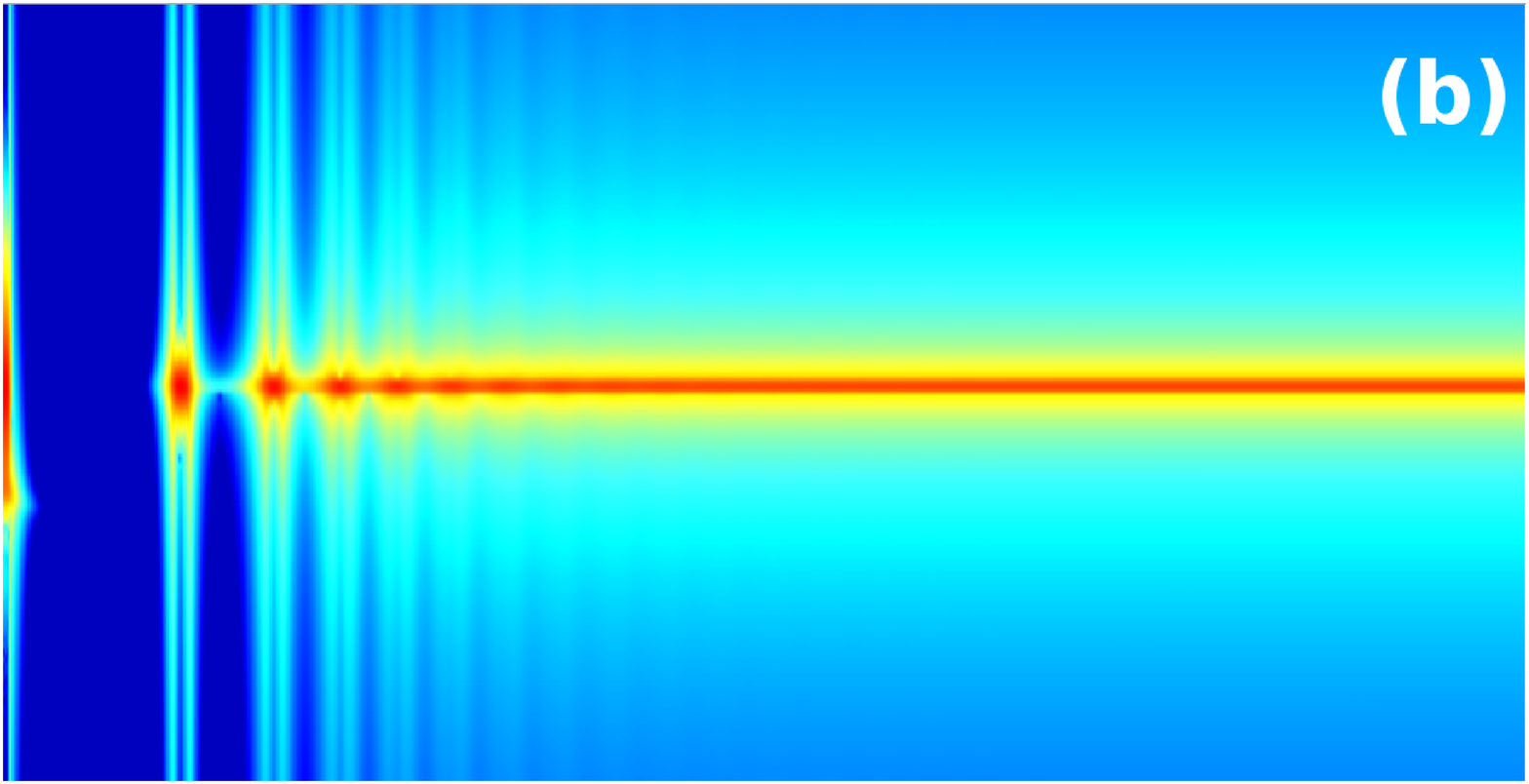} \\
  \includegraphics[height=1.9cm]{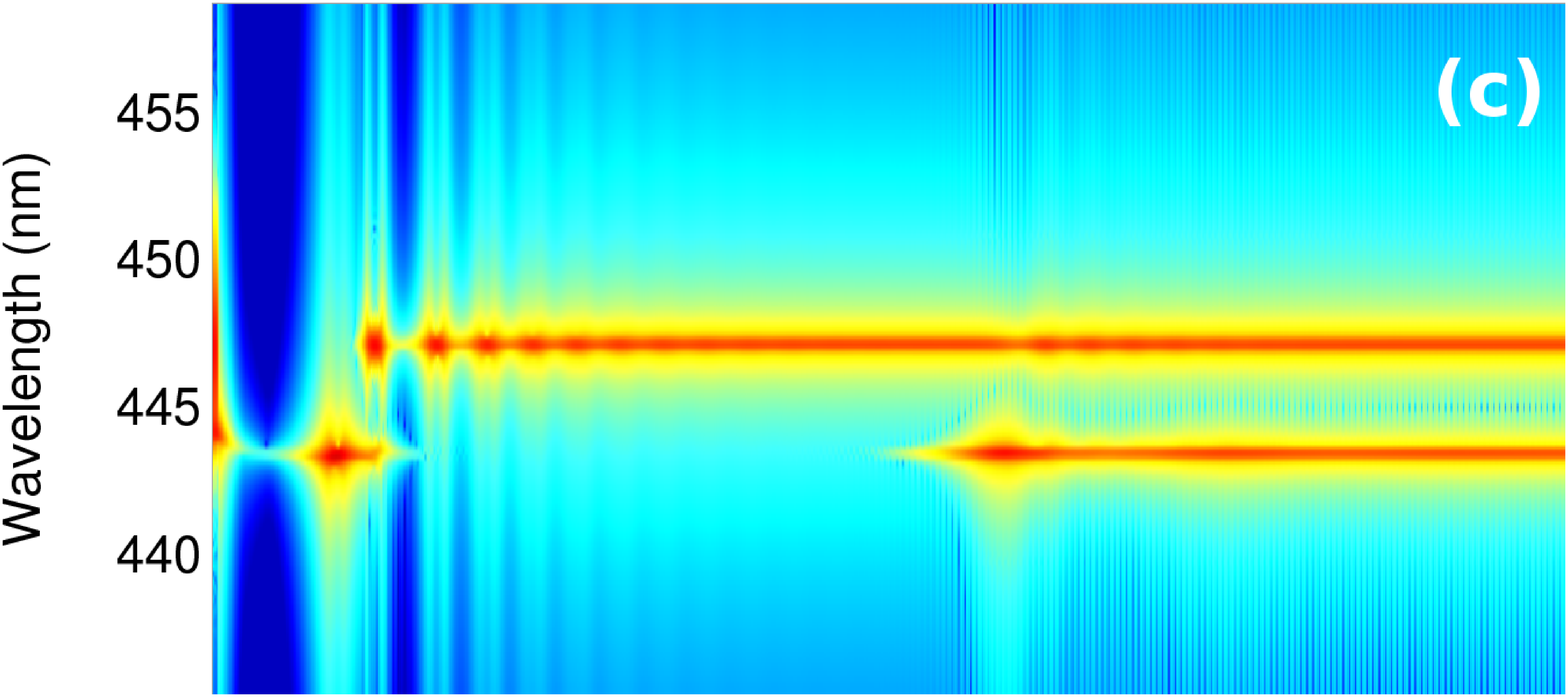} 
  \includegraphics[height=1.9cm]{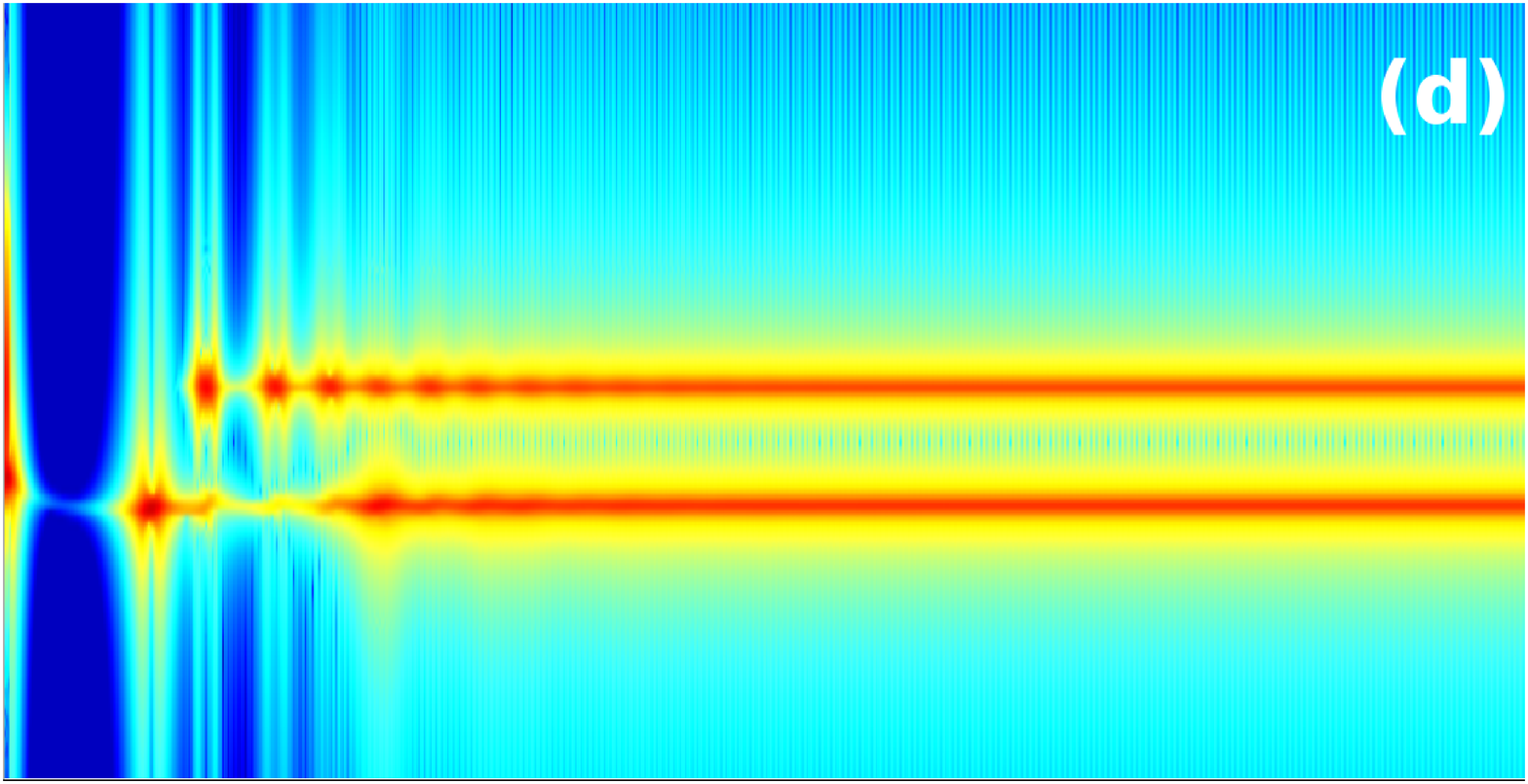} \\
  \includegraphics[height=2.43cm]{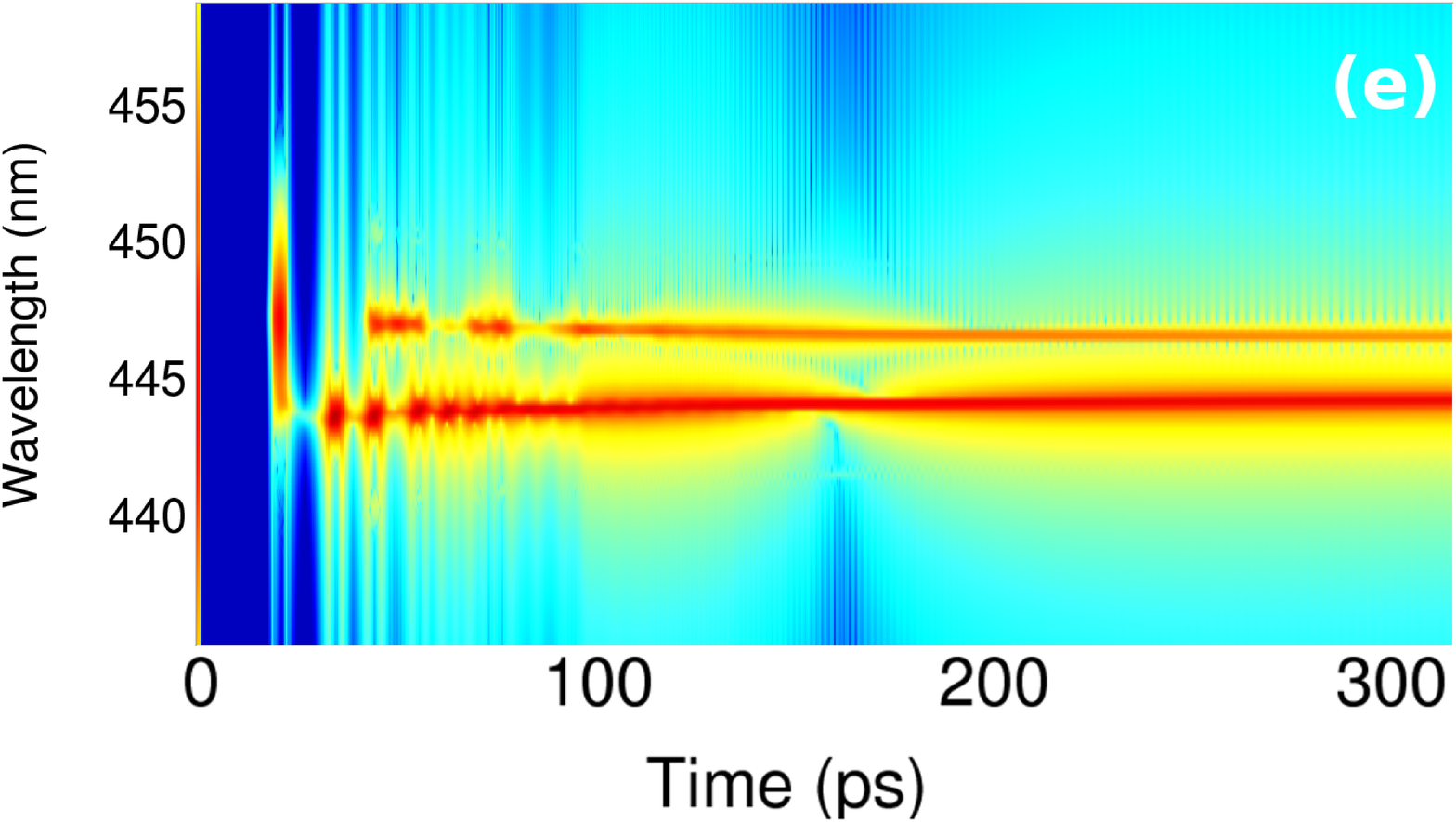} 
  \includegraphics[height=2.43cm]{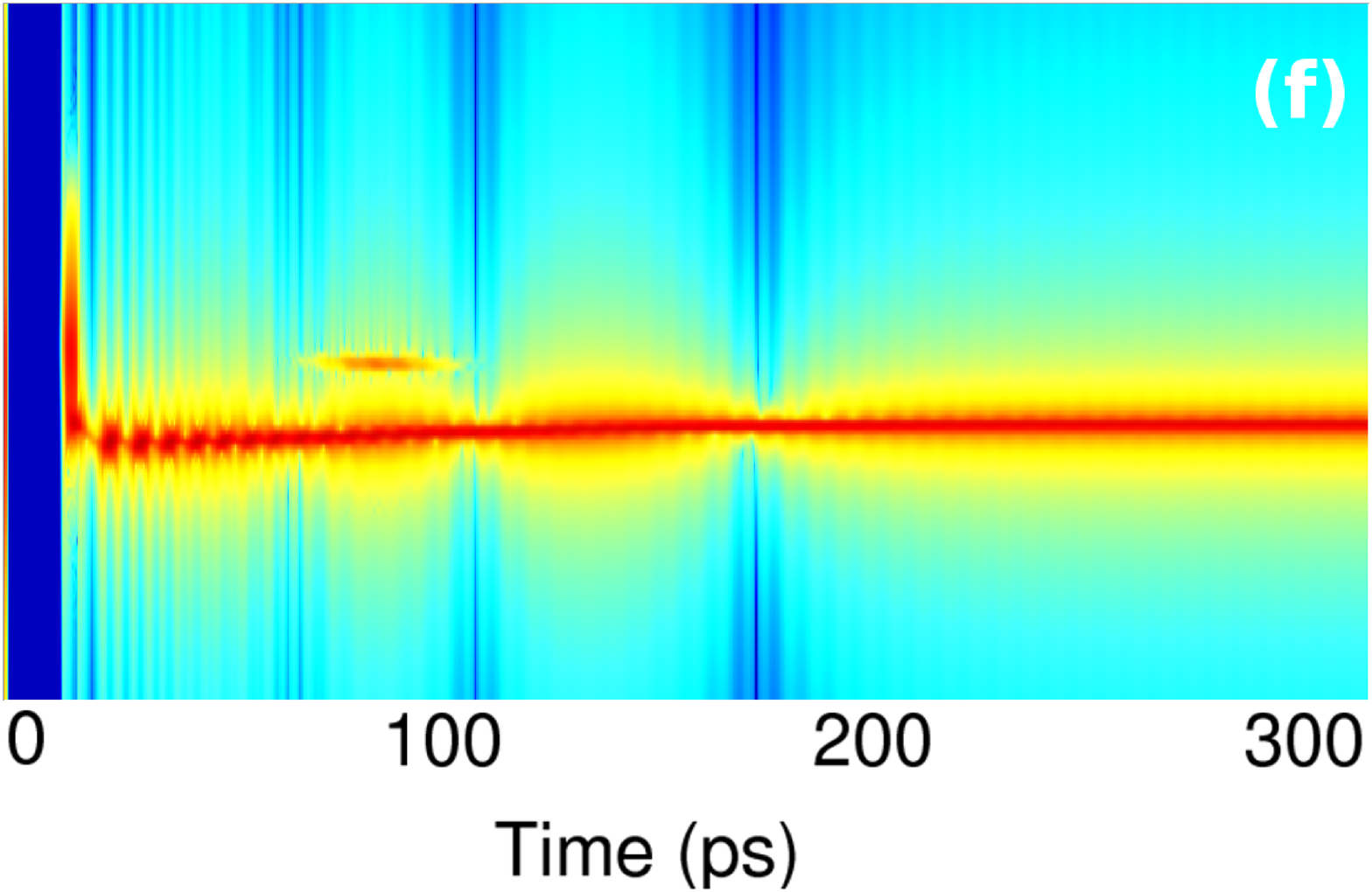} \\
\caption{\label{fig:fig6} (Color online)
  Spectrograms of output intensity (color on a log scale ranging from $10-10^8$)
  of a 1D random laser.
  (a) $P_r =  0.24$ ns$^{-1}$, single-mode lasing. 
  (b) $P_r =  0.29$ ns$^{-1}$, just below the lasing threshold of the second mode. 
  The second mode appears in the transient regime.
  (c--e) $P_r=0.30$, 0.31, 0.43 ns$^{-1}$, multimode lasing.
  (f) $P_r=0.47$ ns$^{-1}$, the first lasing mode is suppressed, though it appears in the transient regime.
}
\end{center}
\end{figure}

Figure \ref{fig:fig6} displays color-coded wavelength power spectra (vertical axis) versus time (horizontal axis) 
for successive values of the pumping rate. 
Spectra are obtained by Fourier transform over a sliding Welch time window of 3.3 ps and a time delay of 0.3 ps.  
Relaxation oscillations are readily seen in all spectrograms at early times as vertical wavelets, which decay rapidly over time. 
Mode competition manifests itself already in Fig.~\ref{fig:fig6}(b) which is just below the lasing threshold of mode 2. 
Mode 2 shows up for a very short duration before the first mode starts oscillating. 
One observes the same phenomenon in Figs. \ref{fig:fig6}(c) and \ref{fig:fig6}(d) but now the second mode comes back after having been suppressed by the onset of the first lasing mode. 
The situation is finally reversed in Fig.~\ref{fig:fig6}(f) where only mode 2 survives, after a brief appearance of mode 1. 
It is interesting to note that the spectral drift of the lasing lines described in Fig.~\ref{fig:fig5}(b) 
occurs in the time domain [Fig. \ref{fig:fig6}(f)]. 
This is a signature of the dynamics of the cross-saturation between lasing modes.

\begin{figure}
\begin{center}
  \includegraphics[height=6cm,angle=270]{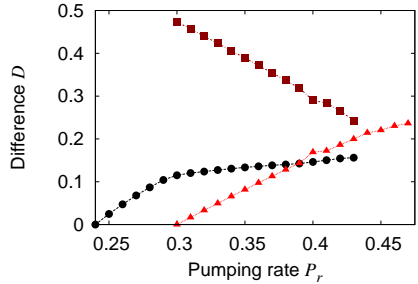}
  \caption{\label{fig:fig7} (Color online)
    Differences $D$ between (circles) lasing mode 1 and lasing mode 1 at threshold,
    (triangles) lasing mode 2 and lasing mode 2 at threshold, and
    (squares) lasing mode 1 and 2 at the same pumping rate $P_r$.
  }
\end{center}
\end{figure}

We consider now the evolution of the spatial distribution of the lasing modes as a function of the pumping rate using the FDTD method.
Figure \ref{fig:fig7} displays the difference $D_{ii}(P_r)$, as defined by Eq. (\ref{Difference}),
between one mode at threshold for lasing ($P_r=P_r^{th}$) and the same lasing mode at pumping rate $P_r$.
The difference increases monotonically with $P_r$, though not at a constant rate.
In this example, the evolution of lasing mode 1 slows down abruptly at the onset of lasing mode 2.
This indicates that self-saturation effects that first dominate the evolution of lasing mode 1 become less efficient at the onset of mode 2.
This is consistent with the evolution of the intensities displayed in Fig. \ref{fig:fig5}(a), which shows that mode 1 slowly
decays when mode 2 appears.
Figure \ref{fig:fig7} also shows that values of the differences become significant when $P_r$ increases.
Hence, the lasing mode distributions depart significantly from the distribution of the associated quasimodes at threshold.
This shows that in open systems like random lasers,
the conventional description of lasing modes in terms of quasimodes of the passive cavity is limited to
values of the pumping rate close to lasing threshold and is not valid for larger pumping rates, a result in agreement with the recent predictions of SALT.

\begin{figure}
\begin{center}
  \includegraphics[height=5cm,angle=270]{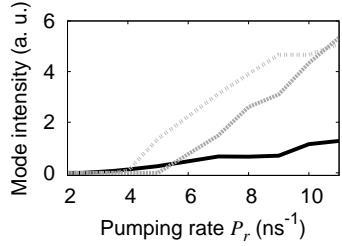}
  \caption{\label{fig:fig8} 
    Intensity vs. pumping rate $P_r$ of the (solid black line) first, (dotted gray line) second,
    and (dashed dark-gray line) third lasing mode of a 2D random laser, size $1\times 1$ $\mu$m$^2$. 
  }
\end{center}
\end{figure}

The previous discussion concerned 1D systems. 
We have also examined 10 different 2D random lasers of size $5\times 5$ $\mu$m$^2$ and 10 different 2D random lasers of size $1\times 1$ $\mu$m$^2$. 
For the smaller size, the density of states is comparable to the 1D systems.
Figure \ref{fig:fig8} shows a representative sample of the evolution of lasing modes for one of those 2D lasers, size $1\times 1$ $\mu$m$^2$.
We observe irregular increase of the mode intensities with increasing pump rate, which again is the signature of mode self-saturation and cross-saturation.
The evolution of lasing mode intensities as a function of $P_r$ is quite similar 
in the more realistic 2D random lasers of larger size.
The difference is that more modes reach the lasing threshold in the larger 2D system, but
similar irregular increase of the mode intensities is observed as in Fig. \ref{fig:fig8}.

\subsection{Nonlinear wave-mixing}

Until now, we have described nonlinear effects involving gain saturation and gain competition, as well as nonlinear refraction. 
In our numerical simulations, in both the single and multimode regimes, we also observe 
the effect of $\chi^{(3)}$ nonlinearities, i.e., frequency-mixing processes including third-harmonic generation and four-wave mixing. 

\begin{figure}
\begin{center}
  \includegraphics[width=4.2cm]{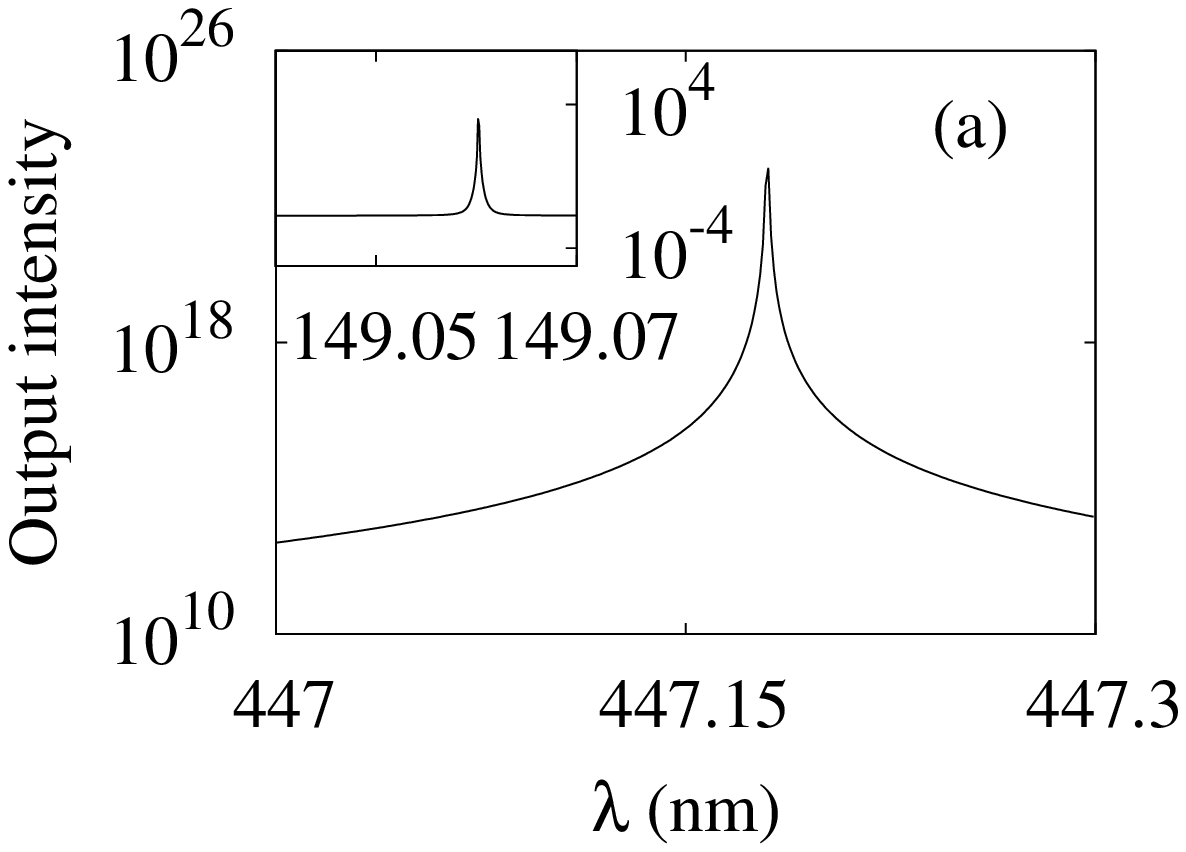}
  \includegraphics[width=4.2cm]{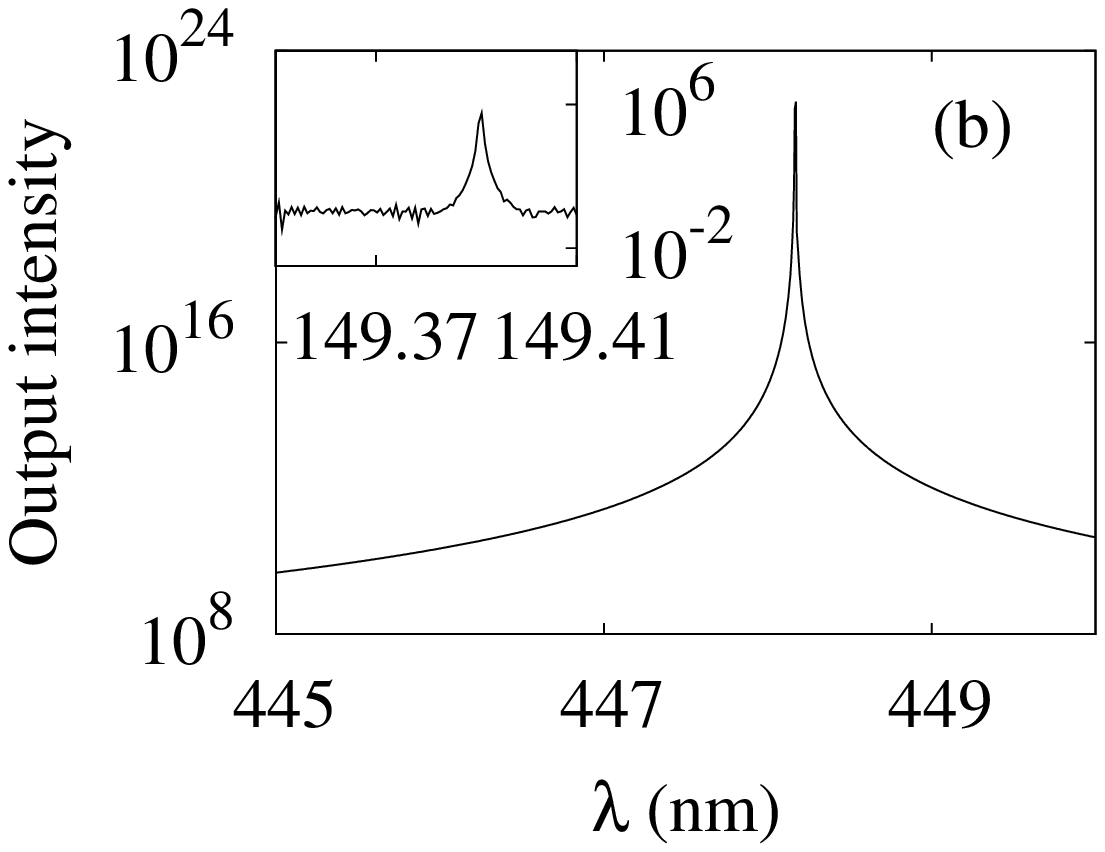}
  \caption{\label{fig:fig9}
    Emission spectrum at the threshold of the first lasing mode and (inset) peak from third harmonic generation for a
    (a) 1D random laser at $P_r = 0.24$ ns$^{-1}$ and
    (b) 2D random laser at $P_r = 2.50$ ns$^{-1}$.
  }
\end{center}
\end{figure}

First, we observe that each lasing mode of wavelength $\lambda$ has a peak at the wavelength $\lambda/3$ associated with it (Fig.~\ref{fig:fig9}). 
We found that third harmonic generation is an efficient process in random lasers, which occurs whatever the value of the pumping rate at or above threshold.

\begin{figure}
\begin{center}
  \includegraphics[width=4.2cm]{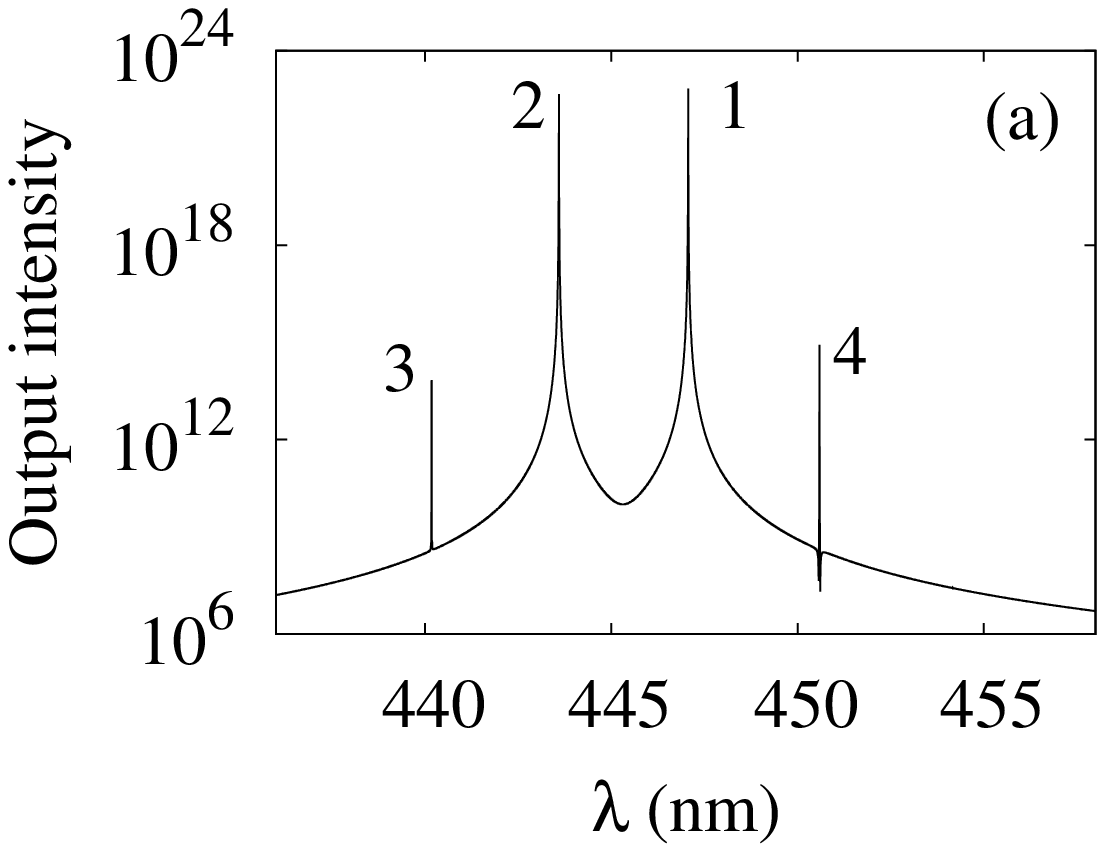}
  \includegraphics[width=4.2cm]{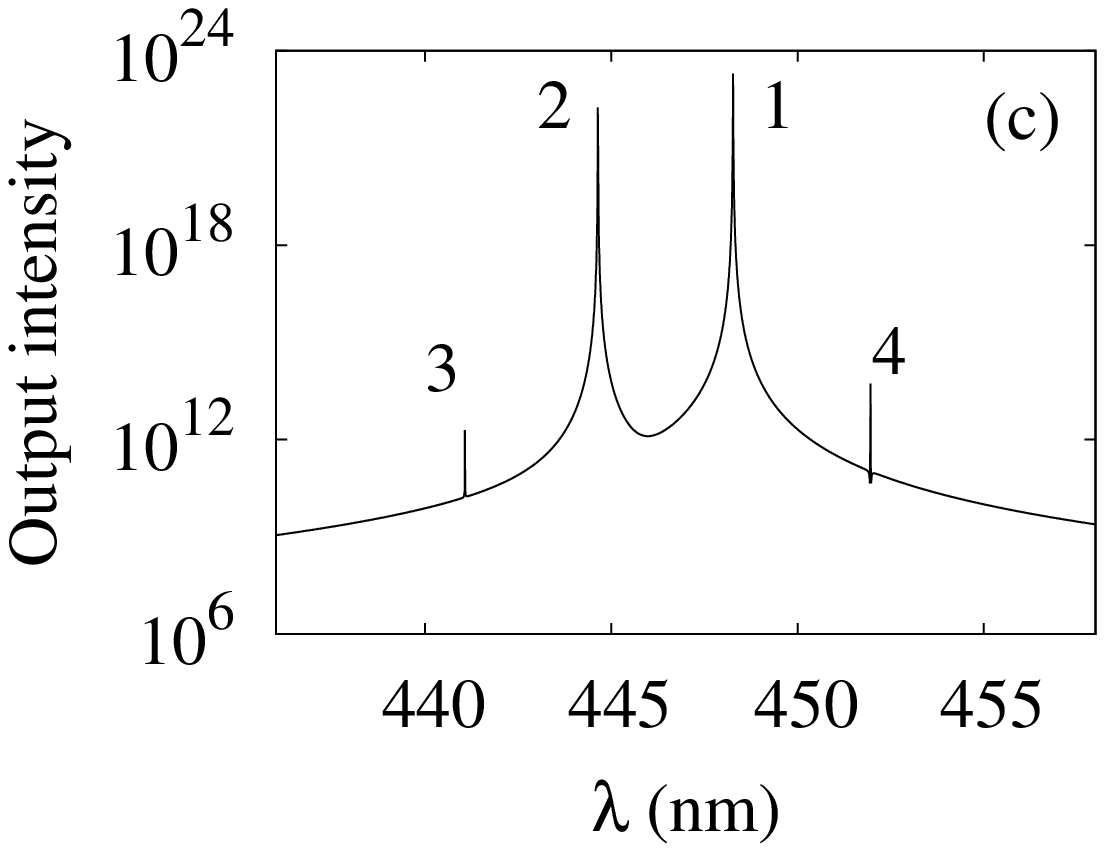}\\
  \includegraphics[width=4.2cm]{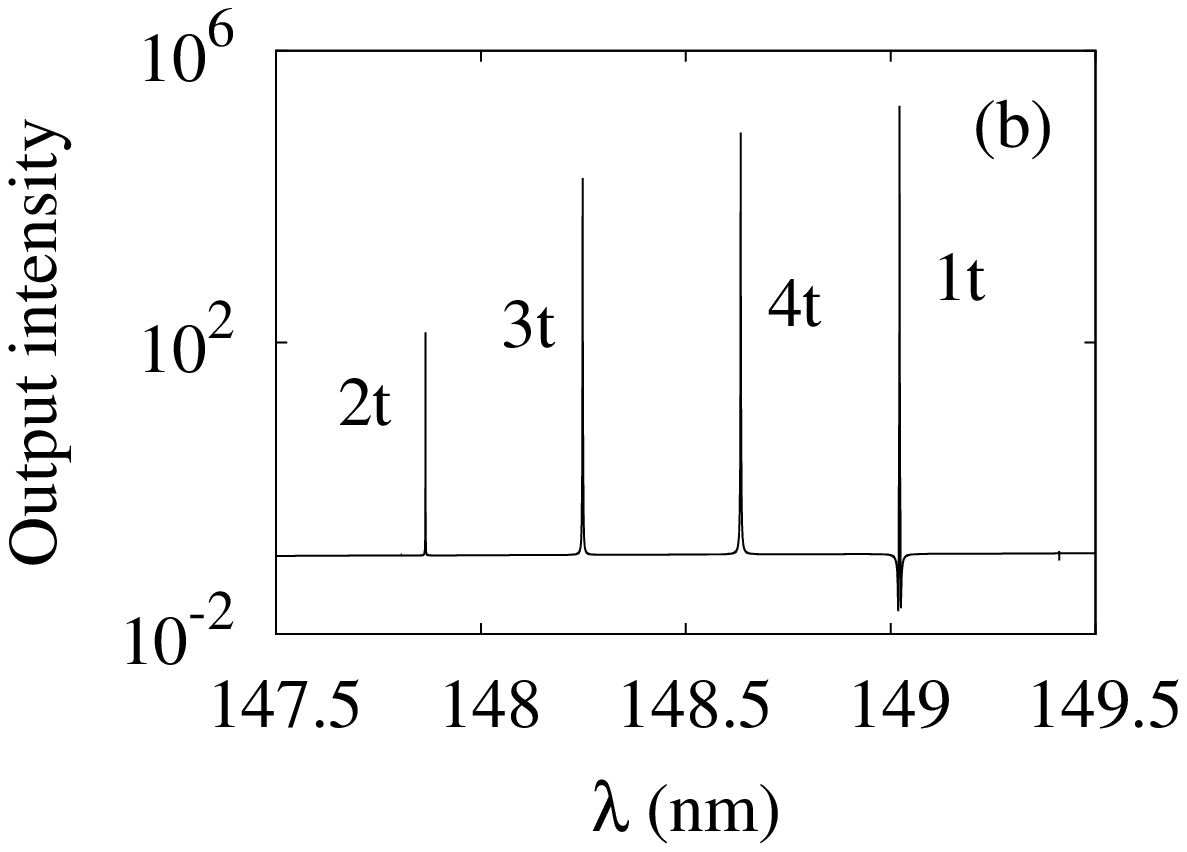}
  \includegraphics[width=4.2cm]{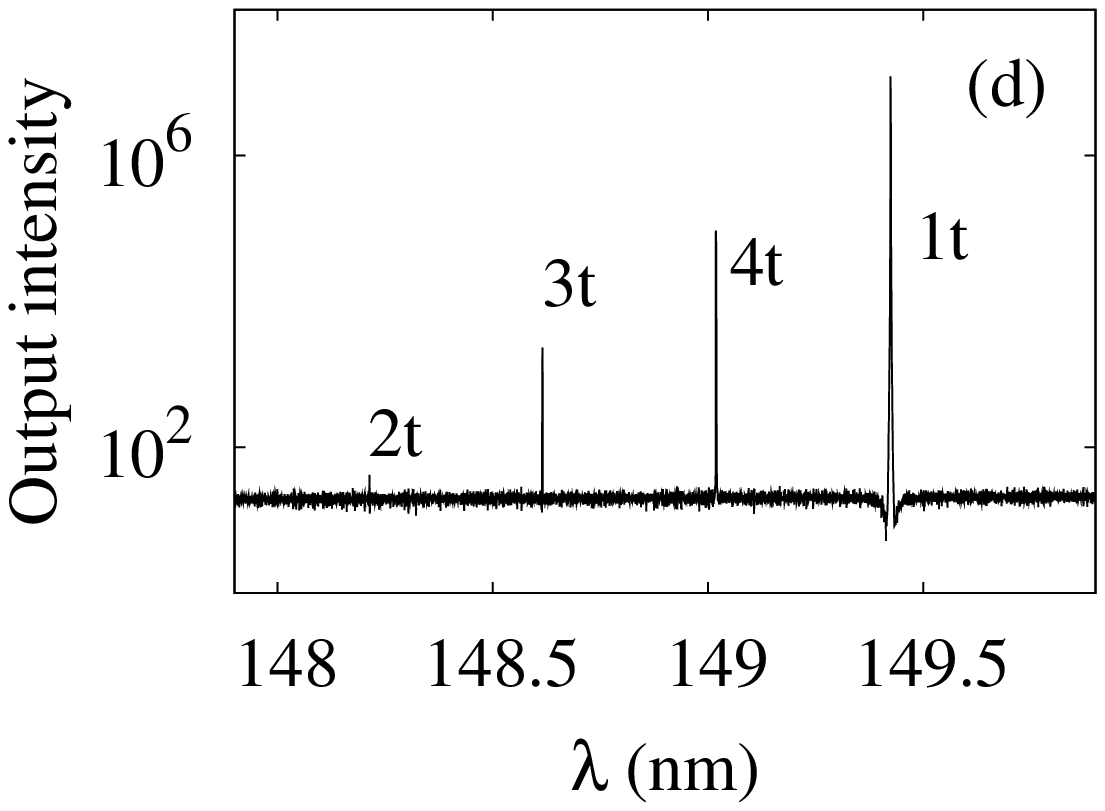}
  \caption{\label{fig:fig10}
    Emission spectrum for a
    (a-b) 1D random laser at $P_r = 0.30$ ns$^{-1}$ and
    (c-d) 2D random laser at $P_r = 3.00$ ns$^{-1}$.
    (a,c) Two lasing modes (labeled 1 and 2) 
    and the peaks resulting from four-wave mixing (labeled 3 and 4). 
    (b,d) Peaks resulting from third harmonic generation (labeled 1t, 2t) and
    sum-frequency generation (labeled 3t, 4t).
  }
\end{center}
\end{figure}

Next, at the onset of a second lasing mode, we observe new peaks in the emission spectrum. 
Examples for 1D and 2D systems are displayed in Figs. \ref{fig:fig10}(a) and \ref{fig:fig10}(c). 
On top of the two lasing modes with wavelengths $\lambda_1$ and $\lambda_2$, 
narrow lateral peaks appear at $\lambda_3$ and $\lambda_4$. 
We attribute the origin of these new lines to four-wave mixing. 
This can be checked as follows
\begin{subequations}
  \label{FWM}
  \begin{eqnarray}
    && 1/\lambda_3-1/\lambda_2=1/\lambda_2-1/\lambda_1 \\
    && 1/\lambda_4-1/\lambda_1=1/\lambda_1-1/\lambda_2
  \end{eqnarray}
\end{subequations}
This signature of four-wave mixing has been verified for all couples of neighboring lasing modes in 1D and 2D lasers.
Normally, four-wave mixing requires not only energy conservation, such as Eqs. (\ref{FWM}), but also phase-matching conditions.
In standard optical homogeneous systems, phase-matching is discussed in terms of wave vectors
\begin{equation}
  \label{HFWM}
  \overrightarrow{k_1}+\overrightarrow{k_2}=\overrightarrow{k_3}+\overrightarrow{k_4}
\end{equation}
Obviously, one cannot associate a unique wavevector to the lasing modes of a random system.
Hence for a random laser, Eq. (\ref{HFWM}) is meaningless.
Nevertheless, the systematic observation of four-wave mixing shows that randomness in random lasers naturally provides random quasi-phase-matching conditions.

Associated with four-wave mixing, we observe again third-harmonic generation and sum-frequency generation involving the two lasing modes. 
This is shown in Figs. \ref{fig:fig10}(b) and \ref{fig:fig10}(d). 
One can check the following equalities
\begin{subequations}
  \label{THG}
  \begin{eqnarray}
    && 1/\lambda_{3t} = 2/\lambda_2 + 1/\lambda_1 \\
    && 1/\lambda_{4t} = 2/\lambda_1 + 1/\lambda_2
  \end{eqnarray}
\end{subequations}

To our knowledge, this is the first report of such nonlinear effects in random lasers.
An important result of this section is that nonlinear wave-mixing effects seem to be promoted by the randomness of such systems.
Similar conclusions for three-wave mixing in random nonlinear materials have been reported in the past \cite{Miller1964,Dewey1975}
and more recently \cite{Morozov2004,Melnikov2004,Baudrier2004,Skipetrov2004}. 
Such studies reported that phase-matching conditions, which can be difficult to achieve in pure and homogeneous materials, can be met more easily in a random structure although conditions are not optimized. 
Our results concerning random lasers confirm that this so-called random quasi-phase-matching \cite{Baudrier2004} works remarkably well in nonlinear random media.

\section{Conclusion\label{sec:conclusion}}

In this paper, we have presented numerical results about random lasing when the external pumping rate is progressively increased above threshold. 
Strong mode competition and significant alteration of the lasing modes due to the large pumping powers that are required to compensate the loss have been observed. 
Moreover nonlinear wave-mixing was shown to take place consistently. 
Such results confirm and extend the modal description of random lasers that has been recently established after the long debate about their nature. 
Our results show that the complexity and the openness of random lasers lead to nonlinear effects that are more pronounced than in conventional lasers.

\section*{Acknowledgments}
  We thank O. Alibart and H. Cao for stimulating discussions.
  This work was supported by the ANR under Grant No. ANR-08-BLAN-0302-01,
  the PACA region, the CG06, and the Groupement de Recherche 3219 MesoImage.
  JA acknowledges support from the Chateaubriand Fellowship.
  This work was performed using HPC resources from GENCI-CINES (Grant 2010-99660).

\appendix
\section{Laser equations\label{app:mectpe}}

We write below Maxwell's equations for a 2D system in the case of transverse magnetic polarization and the population equations of the four-level atomic system,
\begin{align}
  & \mu_0 \partial H_x/\partial t = - \partial E_z /\partial y \label{Maxwell}\\
  & \mu_0 \partial H_y/\partial t = \partial E_z /\partial x \nonumber\\
  & \epsilon_i \epsilon_0 \partial E_z/\partial t + \partial P/\partial t= \partial H_y /\partial x - \partial H_x /\partial y,\nonumber
\end{align}
where $\epsilon_i=n_i^2$\quad$i=1,2$, $\epsilon_0$ is the electric permittivity and $\mu_0$ the magnetic permeability of vacuum.
$P$ is the polarization density, which acts as a source in Maxwell's equations.
The time evolution of the four-level atomic system is described by population equations \cite{Siegman1986}.
\begin{align}  
  & dN_1/{dt} = {N_2}/{\tau_{21}} - P_rN_1\label{Populations}\\
  & d{N_2}/{dt} = {N_3}/{\tau_{32}} - {N_2}/{\tau_{21}}-({E_z}/{\hbar\omega_a}){dP}/{dt}\nonumber\\
  & d{N_3}/{dt} = {N_4}/{\tau_{43}} - {N_3}/{\tau_{32}}+({E_z}/{\hbar\omega_a}){dP}/{dt}\nonumber\\
  & d{N_4}/{dt} = -{N_4}/{\tau_{43}} + P_rN_1,\nonumber
\end{align}
$N_i$ is the population density in level $i,\quad i=1$ to 4.
The electrons in the ground level 1 are transferred to the upper level 4 by an external pump at a fixed rate $P_r$.
Electrons in level 4 flow downward to level 3 by means of nonradiative decay processes with a characteristic time $\tau_{43}$.
This time is very short so that the electrons excited in level 4 quickly populate level 3.
The intermediate levels 3 and 2 are the upper and lower levels, respectively, of the laser transition.
The decay rate downward from level 3 is $\tau_{32}$.
Stimulated transitions due to the electromagnetic field take place between these two levels.
Electrons then decay nonradiatively from level 2 to level 1 with a characteristic time $\tau_{21}$.
The stimulated transition rate is given as $({E_z}/{\hbar\omega_a}){dP}/{dt}$,
where $\omega_a=(E_3-E_2)/{\hbar}$  is the transition frequency between levels $2$ and $3$.
The quantities $E_z$, $P$ and $N_i$ depend on the time $t$ but also on the position $\overrightarrow{r}$ in the system.
Eventually, the polarization obeys the equation
\begin{equation}
  \label{Polarization}
        {d^2P}/{dt^2}+\Delta \omega_a {dP}/{dt}+\omega_a^2 P=\kappa \Delta N E_z,
\end{equation}
$\Delta N = N_2-N_3$ is the population difference density between the populations in the lower and upper levels of the atomic transition.

Amplification takes place when the external pumping mechanism produces an inverted population difference $\Delta N < 0 $.
The linewidth of the atomic transition is $\Delta \omega_a = {1}/{\tau_{32}} + {2}/{T_2} = {1}/{T_1} + {2}/{T_2}$,
where we have used the usual notation $T_1$  for $\tau_{32}$.
The collision time  $T_2$ is usually much smaller than the lifetime $T_1$.
The constant $\kappa$  is given by
\begin{equation}
  \label{Constant}
  \kappa = {6\pi \epsilon_0 c^3}/{\omega_a^2 T_1}.
\end{equation}
We have chosen  $\omega_a / {2\pi} = 4.215\times 10^{15}$ Hz corresponding to a gain curve centered at $\lambda_a=446.9$ nm
and the following values $T_1=100$ ps and $T_2=20$ fs.
Hence, the gain curve has a spectral width $\Delta\lambda_a=11$ nm.

\section{Quasimode contributions to lasing mode suppression\label{app:qmtlm}}

\begin{figure}
\begin{center}
  \includegraphics[height=8.6cm,angle=270]{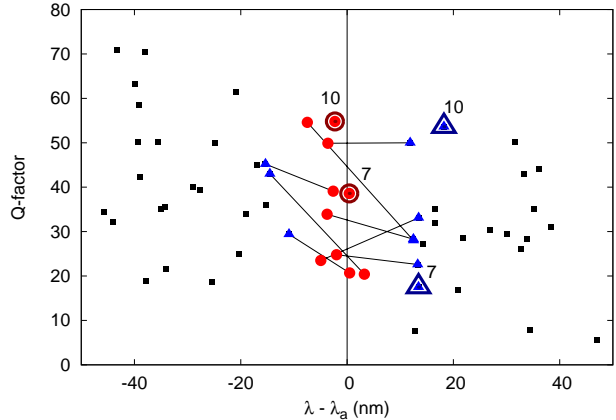}
  \caption{\label{fig:fig11} (Color online)
    Quasimode $Q$-factors and spectral distance from the gain center wavelength $\lambda_a$ for 10 realizations
    of passive random structures in 1D.
    Quasimodes correspond to the first lasing mode (red circles) and second lasing mode (blue triangles)
    found without mode interaction (TM); the first and second modes for the same realization of disorder are connected with a line.
    With mode interaction (FDTD), the two cases in which the first lasing mode strongly suppresses the
    second lasing mode are marked as 7 and 10.    
  }
\end{center}
\end{figure}

Figure \ref{fig:fig11} displays the quasimode quality factors ($Q$-factors) and the spectral distance from the gain center wavelength
for 10 realizations of passive 1D random structures.
The $Q$-factor of a quasimode is defined by $Q=\nu/\delta\nu$ where $\nu$ is the quasimode frequency and
$\delta\nu$ its width equal to the reciprocal of its lifetime $\tau$.
One observes that the quasimodes associated with the first lasing modes are all located near the wavelength
corresponding to the maximum of the gain curve while the quasimodes associated to the second lasing modes are all farther from $\lambda_a$.
In this case, for a quasimode likely to be lasing, the proximity to $\lambda_a$ is more important than the value of the $Q$-factor.

In Fig. \ref{fig:fig4}, it was seen that two realizations of disorder for 1D random lasers exhibited an extremely robust single-mode lasing regime.
The quasimode properties of those two realizations (7 and 10) are shown explicitly in Fig. \ref{fig:fig11}.
For both realizations, the quasimode associated with the first lasing mode is nearly coincident with $\lambda_a$.
For realization 7, the quasimode associated with the second lasing mode is away from $\lambda_a$ and its $Q$-factor is half of the first.
For realization 10, the $Q$-factors of both quasimodes are the same but the second quasimode is farther from the gain center.
Thus, both wavelength and $Q$-factor play a role in the suppression of the second lasing mode.
 
\bibliographystyle{osajnl}

\end{document}